\documentclass[preprint,12pt, 3p]{elsarticle}

\usepackage{amssymb}

\usepackage{bm}
\usepackage{graphicx}
\usepackage{amsmath, amsthm}
\usepackage[usenames]{color}
\usepackage[normalem]{ulem}

\usepackage{amsfonts,amscd}
\usepackage{mathrsfs}
\usepackage{eucal} 	 	
\usepackage{verbatim}      	
\usepackage{makeidx}       	
\usepackage{braket}

\usepackage{natbib}
\usepackage{bibentry}
\biboptions{sort&compress}
\usepackage{hyperref}

\journal{Annals of Physics}

\begin{document}
	
	\begin{frontmatter}
		
		
		
		\title{Gauge Invariant Propagators and States in Quantum
			Electrodynamics}
		
		
		\author[inst1,inst3]{Jordan Wilson-Gerow}
		
		\author[inst1,inst2,inst3]{P.C.E. Stamp\corref{cor1}}
		\ead{stamp@phas.ubc.ca}
		\cortext[cor1]{Corresponding author.}

		\affiliation[inst1]{organization={Department of Physics and Astronomy,
				University
				of British Columbia},
			addressline={6224 Agricultural Rd.},
			city={Vancouver},
			postcode={V6T 1Z1},
			state={B.C.},
			country={Canada}}
		\affiliation[inst2]{organization={Pacific
				Institute of Theoretical Physics, University of British
				Columbia},
			addressline={6224 Agricultural Rd.},
			city={Vancouver},
			postcode={V6T 1Z1},
			state={B.C.},
			country={Canada}}
		\affiliation[inst3]{organization={TAPIR, Walter Burke Institute for
				Theoretical Physics},
			addressline={MC 350-17,	California Institute of Technology},
			city={Pasadena},
			postcode={91125},
			state={CA},
			country={USA}}
		
		\begin{abstract}
			We study combined matter/gauge field gauge invariant states in
			terms of data living on the boundary of gauge invariant path-integrals. To get
			concrete results, this is done for scalar and spinor QED, for both
			`time-slice', and `causal diamond' boundaries. We discuss both the standard
			case where the gauge field falls off to zero at the spatial boundaries, and
			the case of `large gauge transformations', where it remains finite at these
			boundaries. The path-integral naturally generates a specific dressing factor
			for states living on time-slices, without fixing any gauge, and we identify a
			universal contribution which depends only on the nature of the boundaries. We
			also derive the analogous dressing for states defined on null infinity,
			showing both its Coulombic parts as well as soft-photon parts.
		\end{abstract}
		
		
		
		
	\end{frontmatter}
	
	
	
	
	
	\section{Introduction}
	\label{sec:intro}

	
	
	Traditionally, both non-relativistic quantum mechanics (QM) and relativistic
	Quantum field theory (QFT) have been formulated in terms of states in Hilbert
	space, upon which operators representing measurements of physical quantities are
	supposed to act at specific times or on some specific hypersurface. However the
	last few decades have seen a growing feeling that one needs to go beyond such a
	framework.
	
	One key motivation for this has come from quantum gravity, where the
	difficulties in defining diffeomorphism-invariant physical quantities
	\cite{bergmann,kuchar,woodard,marolf06} have led to approaches in which states
	are defined in terms of information residing on boundaries
	\cite{hartle88,hartle94,oeckl,rovelliB}. Path integrals can be used to define
	ground-state wave-functions for different kinds of boundary \cite{hawkingPI} or
	even for spacetimes with no boundaries \cite{hartleH,halliwell}. Much of modern
	quantum cosmology is also formulated using path integrals \cite{Qcosmo}.
	
	There have also been extensive efforts to look for generalizations of QM in
	which, eg., the superposition principle breaks down \cite{kibble89,weinberg89};
	in recent years these have focused on the possible role of gravity in
	engineering this breakdown \cite{kibbleQG,penrose96,bassi14,stampCWL,CWL2,CWL3}.
	In all of these developments, alternative definitions of quantum states have
	been sought.
	
	That path integrals provide a more general formulation of QM states has been
	known for a long time \cite{morette82}, and their usefulness has been seen in,
	eg., the Aharonov-Bohm effect \cite{AhB57}, and in interaction-free measurements
	\cite{no-int}. In such examples, the time evolution of a quantum system ${\cal
		S}$ depends both on what can happen along the paths ${\cal S}$ does follow, and
	also those paths it does {\it not} follow. In a path integral formulation this
	seems quite natural - but not if we deal entirely with the wave-function
	$\langle {\bf r}|\Psi_S(t) \rangle$, which is zero in regions where no paths are
	followed.
	
	In more recent years path integrals have also been used to deal with
	topological field theories, and in describing states with fractional statistics
	for many-particle systems \cite{fracStat,wu84,stone}; in these systems,
	boundaries play a key role in defining both the ground and excited states.
	
	What almost all of these discussions have in common is (i) their emphasis on
	the role of boundaries in defining quantum states; (ii) the use of the path
	integral in giving this definition; and (iii) the presence of gauge fields. This
	last feature adds a further complication, since one would like to be able to
	define gauge invariant quantities.
	
	A key motivating factor here has been that for general spacetimes, path
	integrals are actually unavoidable. For any non-trivial spacetime in which there
	exist achronal regions, one must employ path integrals to handle the dynamics of
	even simple particles \cite{hartle88,hartle94,thorneCTC,visser}. In these
	situations, a conventional Hamiltonian framework is no longer applicable, and
	simple Hamiltonian evolution is undefined, whereas path integrals can still
	compute transition amplitudes/probabilities.
	
	If one is prepared to accept the existence of non-trivial topologies in
	quantum gravity, then any amplitude must involve sums over them. Dealing with
	such sums has been a major theme in research since the 1980's, invoked to solve
	key problems such as the cosmological constant problem
	\cite{coleman91,hebeker18} and the black hole information loss problem~\cite[see
	eg.][]{penington2020, almheiri2020, vanraamsdonk2020}. In these studies, the
	gravitational path-integral inevitably describes a structure more general than
	just ``wave function \& Hamiltonian'' quantum
	mechanics~\cite{marolf2020,giddings2020}.
	
	For all these reasons, it is clear that a `path-integral first' formulation
	of QM and QFT, in which boundaries play a key role, is desirable.
	
	Much of the work in this area has been quite abstract, and has tried to deal
	with general kinds of boundary and boundary information. In the work of Hartle
	and Hawking \cite{hartleH}, states are defined using path integrals in some
	general spacetime regions $\mathcal{V}$, bounded by some hypersurface
	$\partial\mathcal{V}$. The states are then non-local wave functionals, over
	configurations specified on all of $\partial\mathcal{V}$. Similar ideas have
	motivated the ``general boundary quantum field theory framework''
	\cite{oeckl,rovelliB}; a field configuration on $\partial\mathcal{V}$ is again
	mapped to an amplitude via a path integral over field configurations in
	$\mathcal{V}$.
	
	To test abstract work of this kind, however, one needs to work out the
	details in specific cases. In studies of quantum gravity this has led to
	considerable discussion - an example is provided by the ongoing debate over the
	validity of both no-boundary and tunneling descriptions of the early universe
	\cite{turok,vilenkin,bojowald}.
	
	The complexity of the issues involved makes it clear that one should try
	first to understand how all of this works in simpler models. This is the
	question addressed in the present paper, which deals with ordinary Quantum
	Electrodynamics (QED). We look at states defined by propagators between
	hypersurfaces in 4-dimensional spacetime, and show how one can naturally define
	both boundary contributions and bulk contributions to these propagators.
	
	In this way we are able to not only recover known results - which now appear
	in a new light - but also derive several quite new results, even in flat
	spacetime, in which the role of boundaries in defining the states becomes very
	clear. QED then acts as a blueprint for the larger project of defining quantum
	states in some sort of boundary quantum field framework. We arrive at the
	conclusion that such a non-local formulation of quantum states is inevitable,
	even for QED, and is more natural than the traditional Hamiltonian formulation
	in terms of local states defined in Hilbert space.
	
	In this study one must immediately confront both conceptual and technical
	issues. These involve both the gauge invariance and the asymptotic properties of
	the states, and arise even in flat spacetime. For this reason, in the rest of
	this introduction we first introduce some of the conceptual and technical
	questions involved, before briefly describing the organization of the paper.

	\subsection{Some Key Questions}
	\label{sec:intro-backG}
	

	In gauge theories, a crucial role is played by constraints, and by the
	requirement of gauge invariance. This was recognized early on by Dirac, as part
	of his efforts to quantize constrained theories \cite{dirac50,dirac64}; he used
	operator representations of the constraints to annihilate physical states. Dirac
	was thereby led \cite{dirac55} to introduce gauge-invariant ``physical states"
	in Quantum Electrodynamics (QED); and the constraints were then the generators
	of the QED gauge transformations.
	
	One can also define gauge-invariant states prepared by a path integral,
	which are of course non-local objects. These states will satisfy the operator
	constraints provided that the action, the measure, and the set of summed paths
	are themselves invariant under the transformations generated by the constraints
	\cite{halliwell}. A good example, already noted above, is provided in quantum
	gravity by the Hartle-Hawking ``no-boundary" wave-function of the universe
	\cite{hartleH}, where one has a Euclidean path-integral over four-dimensional
	metrics. This state then satisfies the Hamiltonian constraint of Einstein
	gravity, in the form of the Wheeler-DeWitt equation \cite{deWitt67c}.
	
	However, in both QED and quantum gravity, one must address the following
	issues:
	
	\vspace{2mm}
	
	(i) Eventually one needs to explicitly address the role of the - often
	complicated - dynamics of real charges (in QED) or masses (in quantum gravity).
	It is often not clear how to separate out the `physical' degrees of freedom from
	unphysical ones - this is particularly true when one deals with rapidly
	accelerating objects, where a discussion on terms of `near field' and `far
	field' zones does not help in making such a separation. One can discuss things
	entirely in terms of asymptotic states \cite{ashtekar,strominger,CQG18}, but
	this is not of much help in dealing with phenomena in bounded regions of
	spacetime.
	
	\vspace{1mm}
	
	(ii) If one is dealing with state superpositions involving a large spatial
	separation of charge or mass, real confusion arises in discussion of what are
	the correct physical variables, or how to test, eg., whether or not the
	gravitational metric field $g^{\mu\nu}(x)$ is quantized
	\cite{vedral17,milburn17,beilok18,belenchia18}. The related question of how to
	properly define notions like decoherence is also unclear, with different results
	being derived for decoherence rates by different authors
	\cite{CQG18,anasHu,onigaW,blencowe,jordanMSc}.
	
	\vspace{1mm}
	
	(iii) While integrating separately over gauge field and matter variables in
	a path integral, one needs to deal properly with both the constraints and the
	gauge redundancy. To deal with the latter one typically uses the Faddeev-Popov
	technique \cite{FP67}. This still leaves the problem of implementing the
	constraints in a manifestly gauge invariant way.
	
	\vspace{2mm}
	
	(iv) As soon as one goes beyond Abelian gauge theory, one runs into problems
	defining states, even using path integrals; in non-Abelian theories, these are
	connected with the Gribov ambiguity \cite{gribov}, and in quantum gravity, with
	functional integration over different metrics. In non-trivial spacetimes,
	including those containing achronal regions
	\cite{hartle88,hartle94,thorneCTC,visser}, the question of how to define quantum
	states for physical systems is completely unresolved.
	
	\vspace{2mm}
	
	These issues are all formal in nature, and we will see that by formulating
	everything in terms of path integrals, once can address them (although in this
	paper we do not try to deal with non-Abelian or gravitational theories). There
	are also physical (as opposed to formal) questions we would like to see
	answered. These include:
	
	(a) What sort of electromagnetic dressing is ``chosen" by states defined
	using path integrals? And how are these states related to the states defined by
	Dirac \cite{dirac55}, Mandelstam \cite{mandelstam62}, and others?
	
	(b) How do the physical states so defined depend on the spacetime boundaries
	and the information specified on them? What about cases involving `large' gauge
	transformations, which also act on these boundaries?
	
	(c) What are the physical degrees of freedom involved in gauge invariant
	spatial superpositions and in entangled states?
	
	(d) What are the implications of results found for QED in the larger
	enterprise of defining states in quantum gravity?
	
	At the end of this paper we will return to these questions. Note that there
	is one other very interesting question we will only comment on briefly in this
	paper. This concerns decoherence and information loss; one would like to know
	how to correctly calculate decoherence rates, and which states we should average
	over to do so. This will be dealt with elsewhere (see also refs. \cite{CQG18}).

	\subsection{Organization of Paper} \label{sec:intro-org}
	
	The paper is organized as follows. In section 2 we consider QED in flat
	spacetime, with boundaries defined by 2 time-slices. We first recall the
	standard definition of quantum states for scalar QED, and then show how to
	derive the form of the gauge-invariant propagator between time slices. We use
	this simple example to highlight the gauge independence of the results,
	demonstrate how the boundary phases emerge without fixing a gauge beforehand,
	and then sketch an eikonal argument for the dressings coming from the remaining
	path-integral. We then introducing the boundary Faddeev-Popov trick, and show
	how it gives the same results. In the remainder of this section we then show to
	do the same for spinor QED.
	
	After this warm-up exercise, in section 3 we look at what happens for a more
	general boundary hypersurface. We derive the propagator between states on the
	future and past regions of a large causal diamond, again for flat space. This
	calculation allows one to see how to generalize to much more general boundaries
	- a natural separation occurs between boundary degrees of freedom and bulk
	degrees of freedom, and it becomes clear how to define the physical variables
	for the system.
	
	Up to this point, all the discussion has been for gauge transformations
	which vanish at infinity. In section 4 we lift this restriction, and extend all
	of the previous results to the case of ``large gauge transformations".  This
	leads to an interesting connection with the soft-photon, large gauge
	transformation, and dressed state literature, and allows us to clarify questions
	about physical decoherence in QED.
	
	Finally, in section 5, we return to the questions and general issues posed
	just above, and show how they can be answered using the framework and results
	given here.
	
	Throughout the paper we will use units in which $\hbar=c=\epsilon_{0}=1$,
	and a $-+++$ metric signature.

	
	
	\section{Gauge Invariant Propagators and States in Quantum Electrodynamics}
	\label{sec:QED}

	
	
	In this section we treat the standard case of Quantum Electrodynamics (QED).
	We will not be deriving any startling new results in this section. Instead we
	will be showing how to derive answers in a new way, and how to re-interpret
	them.
	
	To avoid needless clutter, we summarize some of the derivations in this
	section - more detail is given in Appendix A. We begin by recalling how states
	have traditionally been defined in QED, and some of the problems associated with
	this. We then move to a path integral description of generalized states defined
	via propagators. This is first discussed for scalar electrodynamics, and then
	for spinor QED. We show how one can derive expressions for the gauge-invariant
	propagator of the combined matter $+$ EM field system between 2 time slices in
	flat spacetime. We show how the results can derived using a ``boundary value
	Faddeev-Popov" method, as well as by more conventional means.

	\subsection{States in Quantum Electrodynamics}
	\label{sec:QED-state}
	
	
	The question of how to define gauge invariant states in QED has a long
	history. Gauss' law, that $\nabla\cdot {\bf E}=\rho$, obviously does not
	completely specify the electric field ${\bf E}(x)$. One can add to the Coulomb
	solution any divergence-free field. In quantum theory the Gauss law operator
	constraint for physical states, viz.,
	\begin{equation}
		\big(\nabla\cdot\hat{\mathbf{E}}-\hat{\rho}\big)|\Psi\rangle=0,
		\label{coulomb}
	\end{equation}
	also has no unique solution.  In the Schr\"{o}dinger picture, in the
	$\hat{A}_{j}(x)$ field value basis, the electric field operator is conjugate to
	$\hat{A}_{j}(x)$, ie.,
	\begin{equation}
		\hat{E}^{j}(x)=i\frac{\delta\,\,\,}{\delta A_{j}(x)}.
	\end{equation}
	
	Many years ago Dirac argued that one should write the wave-function for a
	static physical electron in the composite form
	\begin{eqnarray}
		\Psi({\bf r}|A^{\mu}) &=& e^{-i \varphi({\bf r}|A^{\mu})} \psi({\bf r})
		\;\; \equiv \hat{U}_C \, \psi({\bf r}) \nonumber \\
		\varphi({\bf r}|A^{\mu}) &=& e\int d^3 r' f^{\mu} ({\bf r} - {\bf r'})
		A_{\mu} ({\bf r'})
		\label{psiD}
	\end{eqnarray}
	where $\psi({\bf r})$ is the wave-function for the `bare' charge.
	
	As Dirac recognized, the phase factor $\varphi({\bf r}|A^{\mu})$ represents
	a ``dressing" function, describing the change in the field induced by the
	charge.
	Dirac chose an intuitively obvious solution for $\varphi({\bf r}|A^{\mu})$,
	writing
	\begin{equation}
		\varphi({\bf r}|A^{\mu}) \;=\; e \nabla^{-2} \partial_j A^j({\bf r})
		\;\; = \;\; - {e \over 4\pi} \int d^3 r' {\nabla_{r'} \cdot {\bf A}({\bf r'})
			\over |{\bf r} - {\bf r'}|}
		\label{phi-D}
	\end{equation}
	so that the dressing was that of a Coulomb field, and automatically
	satisfied (\ref{coulomb}).
	
	One can always add some divergence-free function to (\ref{phi-D}), to get a
	quite different form; a good example is provided by the  `Mandelstam string'
	solution \cite{mandelstam62}, viz.,
	\begin{equation}
		\varphi({\bf r}|A^{\mu}) \;=\; e \int^{\bf r}_{\Gamma} dz^{\mu}
		A_{\mu}(z)
		\label{phi-M}
	\end{equation}
	where $\Gamma$ is some spacelike path terminating at the point ${\bf r}$;
	solutions of this form have attracted interest in various contexts
	\cite{shabanov,giddings}.

	So far this seems straightforward. Suppose however we consider a simple
	superposition of the position eigenstates $|\textbf{x}_{1}\rangle$ and
	$|\textbf{x}_{2}\rangle$ for the charge. The bare particle wave-function is
	simply
	$\psi_{12}(\textbf{r})=\tfrac{1}{\sqrt{2}}(\delta(\textbf{r}-\textbf{x}_{1})+\delta(\textbf{r}-\textbf{x}_{2}))$.
	Dirac's physical-state wave-function is then of the form
	\begin{equation}
		\hat{U}_{C}\,\psi_{12}(\textbf{r}) =
		\frac{1}{\sqrt{2}}\big(e^{-i\varphi(\textbf{x}_{1}|A^{\mu})}\delta(\textbf{r}-\textbf{x}_{1})+e^{-i\varphi(\textbf{x}_{2}|A^{\mu})}\delta(\textbf{r}-\textbf{x}_{2})\big).
	\end{equation}
	This state is not an eigenstate of the longitudinal electric field
	operator, and no longer has a well defined electric field, but instead a
	superposition of Coulomb fields centred on ${\bf x}_{1}$ and ${\bf x}_{2}$.
	
	These ambiguities are compounded in the general case of a moving charge.
	Classically, the Li\'{e}nard-Wiechert solution, which makes explicit reference
	to the trajectory of the moving charge, can still be used. However, quantum
	mechanically one would describe a moving charge using a wave-packet peaked on a
	particular momentum and the Dirac prescription would then generate a continuous
	superposition of the various Coulomb fields.  We clearly cannot isolate out a
	specific physical Coulomb field -- we would have no idea which one to use! The
	superposition of Coulomb fields in no way represents the expected
	Li\'{e}nard-Wiechert field.  Moreover, one could add different divergence-free
	functions into the dressing operator acting on each branch of the
	superposition---the constraint equation alone cannot tell us which is the
	correct/physical dressing for charges, because its solutions are not unique.
	
	Once we sum over multiple paths for a charge, in QED, we must clearly sum
	over multiple field configurations, some of which will involve radiation, others
	not - the problem seems hopeless. Nevertheless we will now see that, in a path
	integral formulation, one can give a unique separation between constrained
	variables and unconstrained radiative variables, which is manifestly
	gauge-invariant.  Along with other results, this separation allows one to see
	quite clearly how the Li\'{e}nard-Wiechert field arises dynamically in QED.

	\subsection{Scalar Quantum Electrodynamics}
	\label{sec:QED-scalar}
	
	
	In scalar QED one considers a single charged particle, without spin, coupled
	to the quantum electromagnetic field. We wish to consider the propagator for
	this system between two time slices (surfaces of constant $t$) in Minkowski
	space.
	
	We will assume here (but not later in the paper) that the gauge field
	$A_{\mu}$, and any possible gauge transformations of it, will vanish
	sufficiently quickly at spatial infinity that surface terms generated by spatial
	integrations by parts can be ignored. In the section 4 we will generalize the
	results to the case of ``large'' gauge transformations.
	
	We begin by considering a non-relativistic particle with position $q$,
	charge $e$, in an external potential $V(q,t)$, and coupled to the
	electromagnetic field $A_{\mu}$. The extension to multiple particles is trivial.
	The action for the system evolving from an initial time $t_{i}$ to a final time
	$t_{f}$ is $S=S_{M}+S_{EM}$,
	where
	\begin{equation}		
		S_{M}[q]=\int_{t_{i}}^{t_{f}}dt\bigg[\frac{1}{2}m\dot{q}^{2}-V(q,t)\bigg]
		\label{S-M}
	\end{equation}
	describes the particle alone, and
	\begin{equation}
		\label{eq:time-sliceaction}		
		S_{EM}[q,A_{\mu}]=\int^{t_{f}}_{t_{i}}d^{4}x\bigg[-\frac{1}{4}F_{\mu\nu}F^{\mu\nu}+A_{\mu}J^{\mu}\bigg]
	\end{equation}
	describes the electromagnetic field along with the coupling to the matter;
	here $F_{\mu\nu}=\partial_{\mu}A_{\nu}-\partial_{\nu}A_{\mu}$ is the
	electromagnetic field tensor, and $J^{0}(x)=e\delta^{3}(x-q(t))$ and
	$J^{j}=e\dot{q}^{j}\delta^{3}(x-q(t))$ are components of the charge current.
	Note here that the current for a charged particle is conserved even when the
	equations of motion are not satisfied, ie. for a general path in the path
	integral.

	
	\begin{figure}
		\begin{center}
			\includegraphics[width=3.2in]{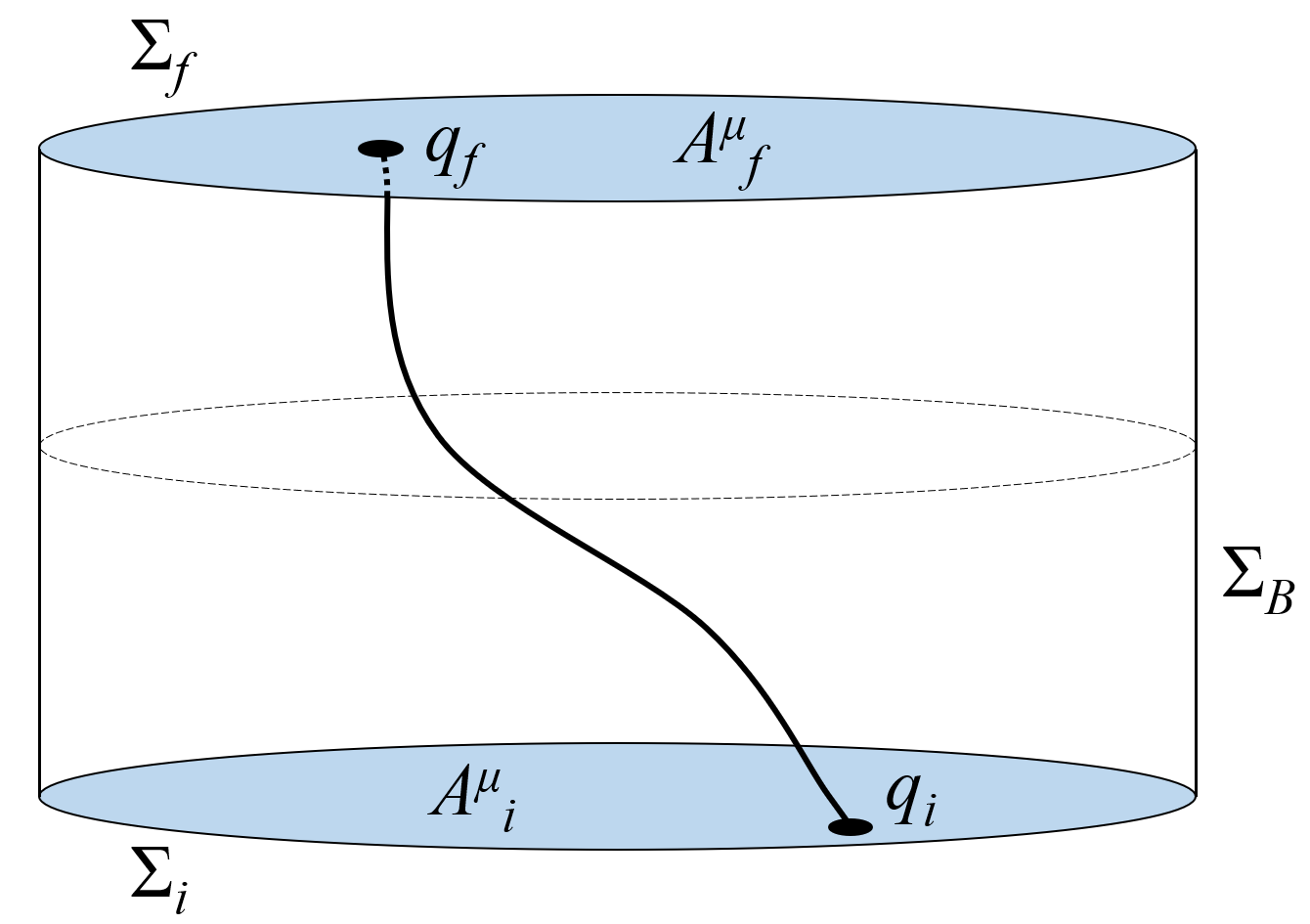}
		\end{center}
		\caption{\label{fig:scalarQED} Depiction of the propagator $K_{fi}$ in
			eqn.
			(\ref{eq:particlepropagator}) and the spacetime through which it
			propagates. The initial configuration is on the time-slice surface $t=t_i$, the
			final configuration on the surface $t=t_f$. Particle paths propagate between
			$q_i$ at $t_i$ and $q_f$ at $t_f$; the gauge field propagates between
			$A^{\mu}_i$ and $A^{\mu}_f$ on these same two time-slices. }
		
	\end{figure}
	

	Under gauge transformation this action transforms by a boundary term. We
	will assume these transformations vanish at spatial infinity, but the
	contribution from the space-like parts of the boundary will not vanish. We
	assume the spacetime region shown in Fig. \ref{fig:scalarQED}. We denote the
	surface of constant time $t=t_{f}$ by $\Sigma_{f}$, and likewise for $t_{i},
	\Sigma_{i}$. An asymptotic timelike cylinder $S^{2}\times\mathbb{R}$ at
	arbitrarily large radius will be denoted by $\Sigma_{\infty}$.
	
	Our path integral is then over field configurations and particle
	trajectories in a region $\mathcal{V}$ bounded by $\partial
	\mathcal{V}=\Sigma_{f}\cup\Sigma_{i}\cup\Sigma_{\infty}$. In this notation,
	under the gauge transformation $A_{\mu}\rightarrow
	A^{\Lambda}=A_{\mu}+\partial_{\mu}\Lambda$, the EM action then acquires a
	boundary term
	\begin{align}
		\delta_{\Lambda} S_{EM} =\int_{\partial V}\,d^{3}x \Lambda
		\,n_{\mu}J^{\mu}&=\int_{\Sigma_{f}}d^{3}x
		\Lambda_{f}J^{0}-\int_{\Sigma_{i}}d^{3}x \Lambda_{i}J^{0} \nonumber \\
		&=e\left(\Lambda_{f}(q_{f})-\Lambda_{i}(q_{i}) \right)
	\end{align}
	
	We will choose to quantize the system on the extended configuration space,
	ie., we consider all configurations of $A_{\mu}(x)$ before quantization rather
	than imposing constraints and gauge conditions at the classical level and
	quantizing the remaining degrees of freedom.
	
	The path integral describing the amplitude for transition between
	configurations $q_{i},A_{\mu\,i}$ and $q_{f},A_{\mu\,f}$ is then
	\begin{eqnarray}
		\label{eq:particlepropagator}
		K_{fi} & \equiv & K(q_{f},A_{\mu\,f};q_{i},A_{\mu\,i}) \nonumber \\
		&=&
		\int^{q_{f}}_{q_{i}}\mathcal{D}q\,e^{iS_{M}}\int^{A_{\mu\,f}}_{{A_{\mu\,i}}}\mathcal{D}A_{\mu}\,e^{iS_{EM}}.
	\end{eqnarray}
	Here and throughout this paper we will absorb field independent constants
	into the path integral measure.
	
	The expression in (\ref{eq:particlepropagator}) is obviously independent of
	the gauge - no choice of gauge enters into it (lest anyone doubt this, we
	demonstrate it explicitly in Appendix A).
	
	Our key aim in this sub-section is to show, by two quite different methods,
	that this expression is completely equivalent to the expression
	\begin{equation}
		\label{eq:finalprop}
		K_{fi} = e^{i\tilde{S}_C}  \int^{q_{f}}_{q_{i}}\mathcal{D}q\,e^{i
			\tilde{S}_{M}}  \int^{\mathcal{A}^j_f}_{\mathcal{A}^j_i} \mathcal{D}
		\mathcal{A}^{j}\,e^{i \tilde{S}_A}.
	\end{equation}
	in which we now employ a different action $\tilde{S}$, given by
	\begin{equation}
		\tilde{S} \;=\; \tilde{S}_M + \tilde{S}_C + \tilde{S}_A
		\label{tilde-S}
	\end{equation}
	with the three new terms defined as:
	
	\vspace{2mm}
	
	(i) The new matter action $\tilde{S}_M$ incorporates a `Coulomb self-energy'
	term, to give
	\begin{equation}
		\tilde{S}_M \;=\; S_M + \tfrac{1}{2} \int_{t_{i}}^{t_{f}}d^{4}x \,
		J^{0}\nabla^{-2}J^{0}
		\label{tildeSM}
	\end{equation}
	with $S_M$ given by (\ref{S-M}) as before; and we note that $\tilde{S}_M$ is
	gauge invariant;
	
	\vspace{2mm}
	
	(ii) The boundary term $\tilde{S}_C$ is a pure phase; we have
	\begin{eqnarray}
		\tilde{S}_C &=& \int_{\partial\mathcal{V}} \sigma \; d^{3}r \,
		J^{0}\nabla^{-2}(\partial_{j}A^{j}) \nonumber \\
		&=& - {e \over 4\pi} \int_{\partial\mathcal{V}} \sigma \; d^{3}r  \,
		\frac{\nabla_{r}\cdot{\bf A}({\bf r})}{|{\bf r} - {\bf q}_{\sigma}|}
		\label{tildeSC}
	\end{eqnarray}
	where $\sigma = \pm$ for future/past boundary time slices (we will
	henceforth omit the $\sigma$, leaving it implicit). The boundary term
	$\tilde{S}_C$
	is just Dirac's Coulomb phase, appearing here on each boundary of the
	propagator.
	
	\vspace{2mm}
	(iii) The `dynamic' part of the EM action $\tilde{S}_A$ is
	\begin{equation}
		\tilde{S}_A \;=\; \tfrac{1}{2}
		\int_{t_{i}}^{t_{f}}d^{4}x\big[-\partial_{\mu}\mathcal{A}^{j}\partial^{\mu}\mathcal{A}_{j}+
		2\mathcal{A}_{j}J^{j}\big]
		\label{tildeSA}
	\end{equation}
	and we see that, like $\tilde{S}_M$, this is gauge invariant. The 3-vectors
	$\mathcal{A}^j_i$ and $\mathcal{A}^j_f$ are the initial and final configurations
	of the `transverse' gauge field $\mathcal{A}^{j}(x)$, itself defined as
	\begin{equation}
		\mathcal{A}_{j}=A_{j}-\partial_{j}\nabla^{-2}(\partial^{k}A_{k})
		\label{tildeA'}
	\end{equation}
	
	This result, as encapsulated in eqns. (\ref{eq:finalprop})-(\ref{tildeA'}),
	may seem surprising - it looks as though the Coulomb gauge has been singled out
	here, and the result is in apparent conflict with ideas, going back to Dirac,
	according to which the Coulomb dressing factor is just one from a large variety
	of possible gauge-equivalent dressing factors which can to describe the quantum
	state of a scalar electron coupled to the EM field.
	
	Nevertheless we will now show, using two different methods, that the result
	(\ref{eq:finalprop}) is not only generally correct, but is gauge-invariant;
	moreover our derivations will avoid any gauge-fixing. We comment on the physical
	meaning of this result both later in this section, and in the conclusions.

	\subsubsection{Method 1: Decomposition of $K_{fi}$}
	\label{sec:method1}
	
	In the first method, one starts from the gauge-invariant expression in
	(\ref{eq:particlepropagator}), and one transforms the action into an equivalent
	form written in terms of new variables, which are themselves determined by the
	constraint equations for the system. The key here is that the form of these
	constraint equations is not determined by any gauge choice - it is instead
	determined by the boundary conditions imposed on our spacetime domain (in this
	case, the boundaries shown in Fig. \ref{fig:scalarQED}).
	
	One begins by making the $U(1)$ gauge transformation: a phase rotation for
	the charged particle, and $A_{\mu} \rightarrow A^{\Lambda}_{\mu}$ for the gauge
	field, as defined above. The gauge-invariant propagator
	(\ref{eq:particlepropagator}) then becomes
	\begin{eqnarray}
		K_{fi}^{\Lambda} &\equiv &
		K^{\Lambda}(q_{f},A_{\mu\,f};q_{i},A_{\mu\,i}) \nonumber \\
		&=&e^{-ie\Lambda_{f}(q_{f})}K(q_{f},A^{\Lambda_{f}}_{\mu\,f};q_{i},A^{\Lambda_{i}}_{\mu\,i})e^{ie\Lambda_{i}(q_{i})}
		\label{K-fi}
	\end{eqnarray}
	where $\Lambda_{i (f)}$ is the gauge parameter on the initial (final) time
	slice.
	
	We know from Hamiltonian dynamics that the Gauss law constraint is the
	generator of gauge transformations. The propagator (\ref{eq:particlepropagator})
	should therefore satisfy Gauss' law as an operator constraint on both
	$\Sigma_{f}$ and $\Sigma_{i}$. Working this through (see Appendix A), one
	obtains the constraint equation for $K_{fi}$ on $\Sigma_f$ as
	\begin{eqnarray}
		0 &=&\bigg[\int d^{3}x\,\partial_{0}\Lambda_{f}\frac{\delta}{\delta
			A_{0\,f}} -i\int
		d^{3}x\Lambda_{f}\bigg(e\delta^{3}(q_{f}-x)-\partial_{j}\hat{E}^{j}\bigg)\bigg]
		\; K_{fi} \qquad
		\label{constraint1}
	\end{eqnarray}
	
	On this surface $\Sigma_{f}$, the functions $\Lambda_{f}$ and
	$\partial_{0}\Lambda_{f}$ are independent, but arbitrary, functions vanishing at
	spatial infinity. As a result, the propagator then satisfies two separate local
	constraint equations
	\begin{align}
		\frac{\delta}{\delta A_{0\,f}} K_{fi} \;&=\; 0 \\
		\bigg(\partial_{j}\hat{E}^{j}-\hat{J}^{0}\bigg) K_{fi} \;&=\; 0.
	\end{align}
	where we see that $K_{fi}$ not only satisfies the Gauss law operator
	constraint, but is also independent of the prescribed data for $A_{0}$. This
	means that we can then freely integrate over the boundary data for $A_{0}$, to
	get
	\begin{equation}
		\label{eq:particlepropagator2}
		K_{fi} \;=\;
		\int^{q_{f}}_{q_{i}}\mathcal{D}q\,e^{iS_{M}}\int\mathcal{D}A_{0}\int^{A_{j\,f}}_{{A_{j\,i}}}\mathcal{D}A_{j}\,e^{iS_{EM}}.
	\end{equation}
	showing that $A_{0}$ is not a true dynamical variable.
	
	That $A_0$ is a redundant variable for this case is of course well known;
	however we would like to emphasize here that:
	
	(i) the boundary data for $A_{0}$ naturally falls out of the expression as a
	consequence of gauge invariance - there is no need for a detour through
	canonical Hamiltonian quantization, or a discussion of the missing conjugate
	momentum $\Pi^{0}$ to see this point. This is because we chose to evolve between
	constant time slices, and the pullback of the 1-form $A_{\mu}dx^{\mu}$ to these
	boundaries is independent of $A_{0}$, making $A_0$ redundant.
	
	(ii) As a corollary to this, we expect that in other boundary geometries,
	the redundant variable will {\it not} be $A_0$. In the next section we will see
	that it is in general just the component of $A_{\mu}$ normal to the boundary,
	just as $A_0$ is the component of $A_{\mu}$ normal to the time slices.
	
	\vspace{2mm}
	
	To complete the derivation of (\ref{eq:finalprop}) we need to write $K_{fi}$
	in terms of the action (\ref{tilde-S}) and the transverse gauge function
	variable $\mathcal{A}_{j}$ defined in (\ref{tildeA'}). This is done by actually
	doing the integration over $A_0$ in (\ref{eq:particlepropagator2}); since this
	derivation is a little long, it appears in Appendix A. The final result is that
	we gave above, in eqn. (\ref{eq:finalprop}).
	
	We see that apart from the natural definition of the transverse vector field
	$\mathcal{A}^{j}$ as the physical EM field variable, we also obtain a Coulomb
	form for the dressing term in $\tilde{S}_C$ in (\ref{tildeSC}), and we see that
	this is {\it not} a consequence of choosing a Coulomb gauge! Instead it arises
	naturally as a boundary term in the path integral expression for $K_{fi}$.

	\subsubsection{Example: Eikonal Approximation}

	Let us briefly discuss what one expects to find from evaluating the
	remaining path-integral over $A_j$ and $q$. We will not attempt to discuss
	detailed examples here. However, the lowest-order eikonal approximation does
	serve to illustrate what one can expect. We give a heuristic treatment here -
	more detail is found in, eg. Fradkin or Fried \cite{CQG18,fradkin,fried}.
	
	In this lowest-order eikonal approximation, fluctuations in the charge
	trajectory about the classical saddle point are neglected in the current.
	Starting from the effective field term $\tilde{S}_A$ in (\ref{tildeSA}), we
	write it as $\tilde{S}_A = \tilde{S}_A^0 + \tilde{S}_A^{int}$, where the
	interaction term is
	\begin{align}
		\tilde{S}_A^{int} \;&=\;
		\int_{t_{i}}^{t_{f}}d^{4}x\mathcal{A}_{j}(x)J^{j}(x)   \nonumber \\
		&=\; e\int_{t_{i}}^{t_{f}}dt\,\dot{q}^{j}(t)\mathcal{A}_{j}(q(t),t)
		\label{S-int}
	\end{align}
	We then expand $q(t)$ as $q(t)=q_{cl}(t)+\delta q(t)$, where the classical
	trajectory $q_{cl}$ is independent of $A_{\mu}$, so that
	\begin{align}
		\tilde{S}_A^{int} \; =&\;
		e\int_{t_{i}}^{t_{f}}dt\,\Big(\dot{q}_{cl}^{j}(t)+\delta
		\dot{q}^{j}(t)\Big) \;
		\sum_{n=0}^{\infty}\frac{1}{n!}\partial_{k_{1}...k_{n}}\mathcal{A}_{j}(q_{cl}(t),t)\delta
		q^{k_{1}}(t)...\delta q^{k_{n}}(t)
	\end{align}
	
	If we were to isolate only the long wavelength parts of $\mathcal{A}_{j}$,
	we could truncate the above derivative expansion at $n=0$; moreover, the high
	frequency trajectory fluctuations would not effectively couple to these long
	wavelength parts of $\mathcal{A}_{j}$ so the term linear in $\delta \dot{q}^{j}$
	would also be negligible.  Discarding these terms is of course only valid for
	the long-wavelength parts of the gauge field, but if we were to simply carry
	this through for the whole field then we effectively perform a lowest order
	eikonal approximation.
	
	The lowest-order contribution for the path-integral in (\ref{eq:finalprop})
	then comes simply from replacing the original interaction term in the action by
	one involving just the classical path of the particle:
	\begin{align}
		S^{eik}_{int} \; &=\;
		e\int_{t_{i}}^{t_{f}}dt\,\,\dot{q}_{cl}^{j}(t)\mathcal{A}_{j}(q_{cl}(t),t)
		\nonumber \\
		&\; \equiv \; \int_{t_{i}}^{t_{f}}d^{4}x\mathcal{A}_{j}(x)J^{j}_{cl}(x).
	\end{align}

	The functional integral for the gauge field coupled to an external classical
	source can be done exactly, and the resulting functional dependence on
	$\mathcal{A}_{j\,f}$ is Gaussian. Assuming the initial state of the gauge field
	is also some Gaussian state, eg. the vacuum, then the Gaussian form remains even
	after using the propagator as a kernel to evolve the initial state. The Gaussian
	functional dependence implies that the out state will generally be a squeezed
	coherent state of the electric field. These linear dynamics cause no squeezing
	of the state, so if the initial state is not squeezed nor will be the out state. 	
	The equivalent statement of this in the canonical quantization framework is
	textbook material, and it is easy to demonstrate because the Heisenberg
	equations of motion are still exactly solvable for a free field coupled linearly
	to a classical source.
	
	We conclude that if the initial state is the vacuum state then, in the
	lowest-order eikonal approximation, the resulting state of the electromagnetic
	field will be a coherent state that is peaked on the classical electric field
	created by the source $J^{j}_{cl}$.
	
	This particular eikonal approximation illustrates very nicely that, aside
	from the universal Coulomb part of the field, coherent dressed states can be
	understood in terms of quantum state preparation \cite{gervais,carney}. As
	laboratory charges are moved around, measured, and probed, they are emitting low
	energy photons and creating a radiative ``dressing''. The eikonal approximation
	also gives a concrete method for computing the long-wavelength parts of the
	dressing resulting from a given state preparation mechanism \cite{CQG18}. It is
	interesting to apply this to real world problems, away from the idealized
	infinite time S-matrix scattering theory, to demonstrate how time-dependent
	dressings emerge dynamically. This is particularly relevant for decoherence via
	soft photons, since the charges are entangled with their radiative dressing.
	
	We can now address one of the questions raised in the introduction.  If one
	considers the dynamics of a charged quantum particle, one finds that the
	expectation value for the long wavelength parts of the electric field operator
	is precisely what is expected from the classical problem, ie., the
	Li\'{e}nard-Wiechert field of a moving charged particle. The dressing can be
	characterized as follows: the resulting state of the electromagnetic field is an
	eigenstate of the longitudinal electric field operator with eigenvalue
	corresponding to the Coulomb field, and a coherent state of the transverse
	electric field which is peaked on a configuration determined by the classical
	limit of the history of the charged particle.  However this field has both
	longitudinal and transverse parts, and by measuring fluctuations one can see
	that they behave quite differently from one another in quantum theory.

	\subsubsection{Method 2: Boundary Faddeev-Popov Trick}
	\label{sec:boundaryFP1}
	
	There is actually no need to use the off-shell current conservation
	constrain we just employed in the demonstration of (\ref{eq:finalprop}). We can
	instead generalize the usual Faddeev-Popov technique to what we will refer to as
	the boundary Faddeev-Popov (bFP) trick; this is similar to a previous technique
	developed in refs. \cite{testa}, but generalized to include quantum matter, and
	to make the gauge independence clear.
	
	We start again from from the manifestly gauge-invariant
	(\ref{eq:particlepropagator}), and, as usual, we multiply the path integral by
	\begin{equation}
		\label{eq:FPintegral}	
		1=\int\mathcal{D}\Lambda\,\Delta[A^{\Lambda}]\delta\big(\mathcal{G}(A^{\Lambda})\big),
	\end{equation}
	where $\Delta[A^{\Lambda}]=|\det\delta_{\Lambda}\mathcal{G}(A^{\Lambda})|$
	is the FP determinant, $\mathcal{G}(A)$ is the gauge fixing function, and again,
	$A_{\mu}^{\Lambda}=A_{\mu}+\partial_{\mu}\Lambda$. In our case the expression
	(\ref{eq:FPintegral}) involves not only integration over gauge transformations
	in the region $\mathcal{V}$, but also over transformations on the boundary time
	slices $\Sigma_{i}\cup\Sigma_{f}$. Transformations residing on the boundaries
	are omitted in textbook applications of the FP trick, where one typically
	considers vacuum generating functionals with no explicit boundaries.
	
	The resulting integral is then
	\begin{eqnarray}
		\label{eq:bFPstep1}
		K_{fi}
		&=&\int\mathcal{D}\Lambda\int^{q_{f}}_{q_{i}}\mathcal{D}q\,e^{iS_{M}} \,
		\int^{A_{\mu\,f}}_{{A_{\mu\,i}}}\mathcal{D}A_{\mu}\,\Delta[A{^\Lambda}]\delta\big(\mathcal{G}(A^{\Lambda})\big)\,e^{iS_{EM}[A]}
		\qquad
	\end{eqnarray}
	Under gauge transformation the FP determinant is gauge invariant, and the
	action transforms by a boundary term
	\begin{equation}\label{eq:bFPstep2}
		S_{EM}[A]=S_{EM}[A^{\Lambda}]-\int_{\partial\mathcal{V}}d^{3}x\Lambda
		J^{0}.
	\end{equation}
	so that the propagator can now be written as
	\begin{align}
		\label{eq:bFPstep3}
		&K_{fi}=\int\mathcal{D}\Lambda\,e^{-i\int_{\partial\mathcal{V}}d^{3}x\Lambda
			J^{0}}\int^{q_{f}}_{q_{i}}\mathcal{D}q\,e^{iS_{M}} \,
		\int^{A^{\Lambda_{f}}_{\mu\,f}}_{{A^{\Lambda_{i}}_{\mu\,i}}}\mathcal{D}A_{\mu}\,\Delta[A]\delta^{\mathcal{V}}\big(\mathcal{G}(A)\big)\delta^{\partial\mathcal{V}}\big(\mathcal{G}(A^{\Lambda})\big)\,e^{iS_{EM}[A]}
	\end{align}
	where we now omit the primes in the notation, and use superscripts
	$(\mathcal{V})$ and $(\partial\mathcal{V})$ to denote quantities evaluated in
	the bulk and the boundary respectively.
	
	Note that both the boundary data for the gauge field, and the delta function
	fixing the gauge on the boundaries, are still dependent on the gauge parameter
	$\Lambda$ -- this of course was not changed by a change of integration
	variables.
	
	In the standard application of the FP trick one would note that there was no
	remaining dependence in the path-integral on $\Lambda$, and the integral over
	the gauge group would simply be divided out as overall normalization; but
	clearly we can't quite do that here. Instead we again use the boundary FP method
	mentioned above. The manipulations are then related to those used in Method 1,
	and we give them in Appendix A. The resulting expression for the propagator is
	\begin{align}
		\label{Kfi-scalar-bFP}
		&K_{fi} \;\;=\; \;
		e^{i\int_{\partial\mathcal{V}}d^{3}x\,\nabla^{-2}(\partial^{k}A_{k})\big[
			J^{0}-i\partial_{j}\frac{\delta}{\delta A_{j}}\big]} \,
		\int^{q_{f}}_{q_{i}}\mathcal{D}q\,e^{i\tilde{S}_{M}}\int^{\tilde{\mathcal{A}}_{j\,f}}_{{\tilde{\mathcal{A}}_{j\,i}}}\mathcal{D}
		\tilde{\mathcal{A}}_{\mu}\,e^{i\tilde{S}_A[\tilde{\mathcal{A}}]} \qquad
	\end{align}
	in which we now write things in terms of the effective actions $\tilde{S}_M$
	and $\tilde{S}_A$, as in (\ref{eq:finalprop}). We can also rewrite this result
	for $K_{fi}$ in the same form as (\ref{eq:finalprop}) - this is also shown in
	the Appendix.
	
	This concludes our analysis of the propagator $K_{fi}$ for scalar QED. We
	stress again that the propagator is gauge-independent, and its form is a simple
	consequence of the gauge invariance of the effective action, rather than being
	imposed {\it a priori}.

	\subsection{Spinor Quantum Electrodynamics}
	\label{sec:diracQED}
	
	We now generalize the above considerations to real QED, where Dirac spinors
	are coupled to the EM field. Again we will begin from a manifestly gauge
	invariant path-integral for $K_{fi}$, and derive the same Coulomb form for the
	dressing. The manipulations are similar to those for scalar electrodynamics, the
	only difference being that the matter field also changes under gauge
	transformation, and the $U(1)$ charge density  in the boundary phase becomes an
	operator.
	
	The gauge invariant path integral representation of the transition
	amplitude, the analogue of (\ref{eq:particlepropagator}) for scalar
	electrodynamics, is
	\begin{equation}
		\label{eq:QEDprop}
		K_{fi}
		=\int_{\psi_{i}}^{\psi_{f}}\mathcal{D}\psi\mathcal{D}\bar{\psi}\int_{A_{\mu\,i}}^{A_{\mu\,f}}\mathcal{D}A_{\mu}\,e^{iS[A,\psi,\bar{\psi}]},
	\end{equation}
	where $\psi,\bar{\psi}$ are Grassmann fields, and the omission of boundary
	data for $\bar{\psi}$ indicates that this variable is to be integrated over on
	the boundary---necessary because the Dirac Lagrangian has a first-order form.
	The action is the QED action with a single Dirac fermion field of charge $e$,
	viz.,
	\begin{eqnarray}
		S[A,\psi,\bar{\psi}] &=&
		\int_{t_{i}}^{t_{f}}d^{4}x\bigg[-\frac{1}{4}F_{\mu\nu}F^{\mu\nu}    - \;
		\bar{\psi}\big(\gamma^{\mu}\partial_{\mu}-ie\gamma^{\mu}A_{\mu}+m\big)\psi\bigg]
		\qquad
	\end{eqnarray}
	
	This action is completely invariant, without need to discard a boundary
	term, under the $U(1)$ gauge transformation
	\begin{eqnarray}
		A_{\mu}&\rightarrow& A_{\mu}+\partial_{\mu}\Lambda \nonumber \\
		\psi &\rightarrow& e^{ie\Lambda}\psi,\hspace{8pt}\bar{\psi}\rightarrow
		e^{-ie\Lambda}\bar{\psi}.
	\end{eqnarray}
	
	One can easily verify that the propagator (\ref{eq:QEDprop}) is gauge
	invariant in the same way done for expression (\ref{eq:particlepropagator}) in
	Appendix A, ie. by transforming its data, undoing the transformation by a change
	of variables in the path integral, and using the invariance of the action.
	
	This gauge invariance of $K_{fi}$ implies that it satisfies the boundary
	equation
	\begin{equation}\label{eq:QEDinvarcondition}
		\int_{\partial\mathcal{V}}d^{3}x\bigg[ie\Lambda\psi\frac{\delta}{\delta\psi}+\partial_{\mu}\Lambda\frac{\delta}{\delta
			A_{\mu}}\bigg]\, K_{fi} \;=\; 0.
	\end{equation}
	
	By explicitly differentiating the path integral we can confirm that the
	functional derivatives are proportional to the conjugate momenta for the fields:
	\begin{equation}
		\label{eq:fermionMomentum}
		\frac{\delta}{\delta\psi_{i,f}}=\pm
		i\hat{\Pi}_{i,f}=\mp\hat{\psi}^{\dagger}_{i,f},
	\end{equation}
	\begin{equation}\label{eq:gaugefieldMomentum}
		\frac{\delta}{\delta A_{j\,i,f}}=\mp i\hat{\Pi}^{j}_{i,f}=\pm i
		\hat{E}^{j}_{i,f}.
	\end{equation}
	
	Together with the expression for the $U(1)$ charge density
	$J^{0}=i\bar{\psi}\gamma^{0}\psi=-\psi\psi^{\dagger}$, and the invariance
	condition (\ref{eq:QEDinvarcondition}), this implies the propagator satisfies
	the operator constraints
	\begin{eqnarray}
		\big(\partial_{j}\hat{E}^{j}-\hat{J}^{0}\big)K(A,\psi) &=& 0 \\
		\frac{\delta}{\delta A_{0}}K(A,\psi) &=& 0,
	\end{eqnarray}
	on both the future and past boundary time slices.
	
	Let us now use the bFP trick again to see how the electric field dressing of
	the states emerges, ie., to see how this constraint is implemented. We again
	insert a gauge fixing function into $K_{fi}$ by multiplying by
	(\ref{eq:FPintegral}), but now we must change variables for both the gauge field
	and the Dirac field if the action is to be invariant:
	\begin{eqnarray}
		K_{fi} &=&
		\int\mathcal{D}\Lambda\int_{\psi^{\Lambda_{i}}_{i}}^{\psi^{\Lambda_{f}}_{f}}\mathcal{D}\psi\mathcal{D}\bar{\psi}
		\int_{A^{\Lambda_{i}}_{\mu\,i}}^{A^{\Lambda_{f}}_{\mu\,f}}\mathcal{D}A_{\mu} \;
		\Delta[A]\delta^{\mathcal{V}}\big(\mathcal{G}(A)\big)\delta^{\partial\mathcal{V}}\big(\mathcal{G}(A^{\Lambda})\big)e^{iS[A,\psi,\bar{\psi}]}
		\qquad
		\label{Kfi-spin}
	\end{eqnarray}
	
	The $A_{0}$ integral can again be done without gauge fixing, and after
	manipulations given in Appendix A, we get
	\begin{equation}
		\label{eq:finalprop2}
		K_{fi} = e^{i\tilde{S}_C}
		\int_{\psi_{i}}^{\psi_{f}}\mathcal{D}\psi\mathcal{D}\bar{\psi} \,e^{i
			\tilde{S}_{M}}  \int^{\mathcal{A}_{j\,f}}_{{\mathcal{A}_{j\,i}}} \mathcal{D}
		\mathcal{A}_{j}\,e^{i \tilde{S}_A}.
	\end{equation}
	which has the same form as (\ref{eq:finalprop}) except that now the matter
	action is
	\begin{equation}
		\tilde{S}_M \;=\; \int_{t_{i}}^{t_{f}}d^{4}x \big[
		-\bar{\psi}\big(\gamma^{\mu}\partial_{\mu}+m\big)\psi +
		\frac{1}{2}J^{0}\nabla^{-2}J^{0} \big]
	\end{equation}
	and the dynamic gauge field action $\tilde{S}_A$ is as before ({\it cf.}
	eqn. (\ref{tildeSA})), except that now the matter current is $J^j =
	ie\bar{\psi}\gamma^{j}\psi$.
	
	We can take the expression in eqn. (\ref{eq:finalprop2}) one step further if
	we explicitly act with the $U(1)$ transformation sitting outside the
	path-integral. This locally rotates the boundary data for the Dirac field by an
	angle which depends on the longitudinal part of the gauge field, to give
	\begin{equation}
		K_{fi}=\int_{e^{-ie\nabla^{-2}\partial^{j}A_{j\,i}}\psi_{i}}^{e^{-ie\nabla^{-2}\partial^{j}A_{j\,f}}\psi_{f}}\mathcal{D}\psi\mathcal{D}\bar{\psi}\int_{\mathcal{A}_{j\,i}}^{\mathcal{A}_{j\,f}}\mathcal{D}\mathcal{A}_{j}\,e^{i\tilde{S}_{M}+i\tilde{S}_{A}}.
	\end{equation}
	for the gauge invariant QED propagator on the extended configuration space,
	in which the limits on the integrals are explicitly given as
	\begin{equation}
		e^{-ie\nabla^{-2}\partial^{j}A_{j}}\psi=\exp\bigg(i\int d^{3}y
		A_{j}(y)\frac{e}{4\pi}\frac{y^{j}-x^{j}}{|y-x|^{3}}\bigg)\psi(x).
	\end{equation}
	
	Thus the gauge invariant propagator dresses every point excitation of the
	Dirac field by a Coulomb electric field sourced by the corresponding point
	charge. This is the key result we wish to emphasize for spinor QED, and it
	parallels that found for scalar QED. Again, as with scalar QED, the transverse
	dressing will be determined dynamically by the remaining integral over gauge
	invariant variables.

	
	\section{Flat spacetime evolution in a causal diamond}
	\label{sec:diamond}
	
	
	Up to now we have dealt with the rather simple problem of QED on a flat
	background, defined between time slices. We now turn to more general kinds of
	boundary and boundary information.
	
	The region we consider is a causal diamond in Minkowski spacetime, where the
	state is fixed on the null boundary hypersurface. Here there is still a natural
	splitting into past and future sections, and so we can define a propagator which
	represents a transition amplitude between states on the past and future null
	cones (which tend to null infinity for an infinitely large diamond).
	
	The derivation proceeds in analogy with the work in the last selection. In
	section 3.A we formulate the problem and show how to transform the effective
	action so as to extract the boundary terms and physical variables. In section
	3.B we then derive the form of the propagator in terms of these variables.
	
	We note here that one quickly encounters a subtlety in the specification of
	boundary data for the path-integral. Because the conjugate momentum on a null
	surface involves a derivative along that surface, specifying the field
	configuration also specifies the conjugate momentum. Giving data on both the
	past and future boundaries then over-specifies the boundary data for the
	classical evolution, and the corresponding interpretation as a quantum amplitude
	is then unclear.
	
	We fix this by specifying ``half'' of the field data in some chosen way
	\cite{colby}. We will assume throughout that it is only the positive frequency
	parts of the field which are specified, ie., we interpret the amplitude in terms
	of coherent states in the Bargmann representation. In the following discussion
	we will avoid making this explicit, so as not to clutter the notation.

	\subsection{Formulation of the Problem}
	\label{sec:diamond-F}
	

	In the time-slice geometry, the variable $A_{0}$ was ultimately unphysical,
	and the remaining variables $A_{j}$ split into purely physical transverse and
	pure gauge longitudinal parts - the transverse part being divergenceless, ie.,
	$\partial^{j}\mathcal{A}_{j}=0$. For more general boundary hypersurfaces, a
	natural idea would be to continue to decompose the field into parts with and
	without divergence. This is not possible, for 2 reasons. First, as before, there
	is still the issue of uniqueness -- given a transverse-longitudinal
	decomposition of the vector field, one can freely add some transverse parts onto
	the longitudinal part and the result still transforms correctly under gauge
	transformation. Second, on null hypersurfaces there is no unique notion of
	divergence -- the induced metric is degenerate, and so there is no unique
	inverse metric with which to define the divergence $h^{jk}\nabla_{j}A_{k}$.
	
	For these reasons we again use a procedure whereby the path integral is used
	to generate a unique decomposition into pure gauge and gauge invariant parts of
	the field.
	
	We recall that for flat time-slice boundaries, the boundary data of the
	component $A_{0}$ was integrated over. The saddle point solution for this
	Gaussian integral, $\tilde{A}_{0}$, determined the g-potential $\Phi$, ie., the
	functional of $A_{j}$ transforming as $\delta_{\Lambda}\Phi=\Lambda$; the pure
	gauge part of $\tilde{A}_{0}$ was the time derivative of $\Phi$, and the
	longitudinal part $A_{j}$ was the gradient of $\Phi$. For more general
	boundaries we will then need to single out the component of $A_{\mu}$ normal to
	the boundary hypersurface. This component will play the same role as $A_{0}$,
	and the pure gauge part of its solution will yield a corresponding g-potential.

	\subsubsection{Coordination specification}
	\label{sec:coord}
	
	To implement these ideas we need to choose coordinates appropriately. We
	pick hypersurface adapted coordinates $x^{\mu}=\{S,y^{k}\}$ such that
	$S=\textrm{const.}$ surfaces foliate the spacetime region $\mathcal{V}$, and the
	boundary hypersurface $\partial\mathcal{V}$ is described by particular values,
	$S=S_{i},S_{f}$. Then, using a coordinate basis it is $A_{S}$ which is the
	component generalizing $A_{0}$, because the pullback of $A_{\mu}dx^{\mu}$ to
	$\partial\mathcal{V}$ will be independent of $A_{S}$.
	
	For a causal diamond in Minkowski spacetime we then need to construct
	coordinates adapted to the boundary null cones.  The coordinates we will use are
	rather intuitive. Consider a sphere of radius $R$ at time $t=0$, and from each
	solid angle send an inwards going radial null ray to the future and to the past.
	These null geodesics will converge at $r=0$ at times $t=R$ and $t=-R$
	respectively, and the surface generated by the null rays is the boundary of our
	causal diamond.

	\begin{figure}
		\begin{center}
			\includegraphics[width=0.4\textwidth]{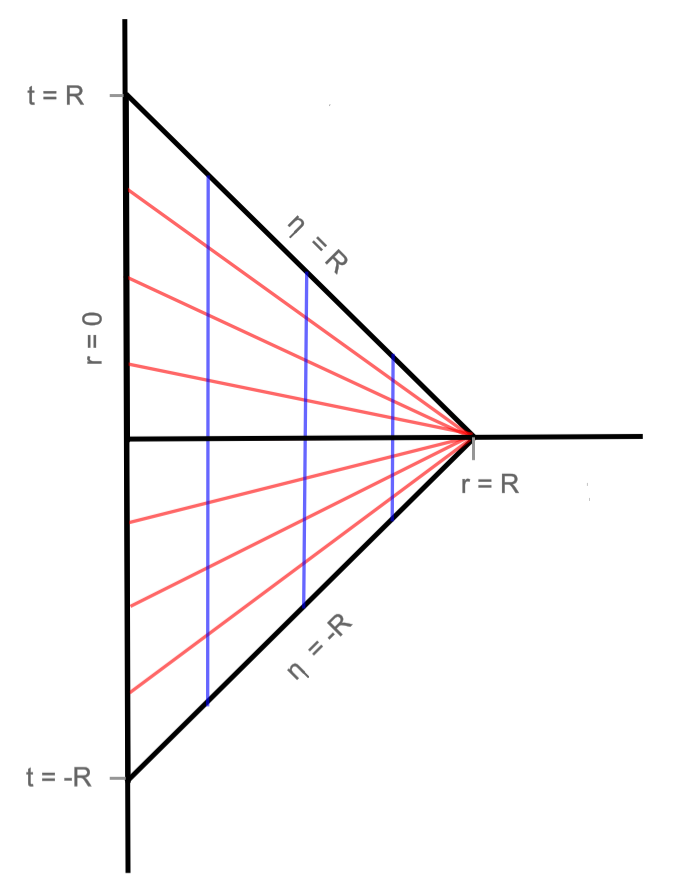}
		\end{center}
		\caption{The $r,\eta$-coordinates. Each point represents a two-sphere of
			radius $r$. This is a standard Minkowski spacetime diagram, not a conformal
			diagram. The blue lines are lines of constant $r$, while the red lines are lines
			of constant $\eta$.}\label{fig:diamondcoords}
	\end{figure}

	To construct coordinates in the interior we again start from the sphere
	$r=R$ at $t=0$, and now send inwards going spacelike rays to the future and
	past. These spacelike rays converge at $r=0$ but at times $t$ dictated by their
	``velocities''. The angles and radii of spheres are still useful coordinates,
	but now we will replace the time coordinate $t$ with a coordinate parameterizing
	the ``velocity'' of each ray.
	
	Each of the rays joining $r=0$ to $r=R$ is described by a solid angle and
	$t,r$ satisfying the simple relation
	\begin{equation}\label{eq:linearrelation}
		t=\eta f(r),
	\end{equation}
	for
	\begin{equation}\label{eq:linearrelation2}
		f(r)=1-\frac{r}{R},
	\end{equation}
	and for some $\eta\in[-R,R]$. From this relation we can quickly verify that
	the surfaces $\eta=\pm R$ are the future and past null boundaries of the causal
	diamond.
	
	Inside the boundary, $\eta$ parameterizes spacelike surfaces and thus serves
	as a useful time coordinate. Thus, as desired, we've found hypersurface adapted
	coordinates where certain values of ``time'' denote the boundary. We can
	straightforwardly compute the metric in these coordinates:
	\begin{eqnarray}
		ds^{2} &=& -f(r)^{2}d\eta^{2}+2\frac{\eta}{R}f(r)d\eta dr  + \;
		\bigg(1-\frac{\eta^{2}}{R^{2}}\bigg)dr^{2}+r^{2}d\Omega^{2} \qquad
	\end{eqnarray}
	where $d\Omega^{2}$ is the standard line element on the unit 2-sphere.
	
	It is clear from this expression that $\eta=0$ is just a standard time slice
	of Minkowski spacetime and that $\eta=\pm R$ are null hypersurfaces. Since
	$f(r)$ vanishes at $r=R$, there is a coordinate singularity. This is obvious
	from Fig. (\ref{fig:diamondcoords}), and indeed several components of the inverse
	metric will diverge as $r\rightarrow R$, 
		\begin{eqnarray}
		&&g^{\eta\eta}=-(1-\frac{\eta^{2}}{R^{2}})\frac{1}{f(r)^{2}},\hspace{20pt} g^{\eta
			r}=\frac{\eta}{R}\frac{1}{f(r)},  \nonumber \\
		&& g^{rr}=1,\hspace{60pt} g^{AB}=r^{-2}q^{AB}  \qquad
	\end{eqnarray}
		where $x^{A}$ are sphere coordinates, and $q^{AB}$ is the inverse metric on
	the unit 2-sphere.
	
	To deal with this we need to recall why we are interested in this geometry.
	Ultimately we wish to take $R$ to be larger than all other length scales so that the
	sphere $r=R$  resembles spatial infinity, and the surfaces $\eta=\pm R$
	resemble null infinity.	 As long as we don't take the strict limit $R\rightarrow
	\infty$, we can still specify data for massive fields on the boundary. The
	boundary considered here then plays a role similar to null infinity, but is not
	obtained via conformal compactification. Timelike worldlines will be able to
	connect all points in the bulk to some point on the boundary. 
	
	Since the electromagnetic field is massless we expect field excitations to
	reach null infinity but we do not expect the same for spatial infinity. For this
	reason we make the assumption that all important quantities will vanish
	sufficiently fast for $r\rightarrow R$, while allowing for finite limits as
	$\eta\rightarrow\pm R$. This physical assumption ensures that the coordinate singularity from $f(r)$ as $r\rightarrow R$ does not actually cause issues during the calculation.	We can see explicitly how these fields vanish by comparing the field components in these coordinates with those in the standard $(t,r)$ coordinates.  For example, for a one-form $w_{\mu}$ we have
	\begin{equation}
		w_{\eta}=f(r)w_{t},
	\end{equation}
	and since physical fields are finite in $(t,r)$ coordinates, $w_{\eta}$ will vanish at least as fast as $f(r)$ as $r\rightarrow R$. 
	
		When it is necessary, we will formally ``blow up'' this surface, ie., excise the sphere $r=R$ from
	the boundary such that limits $r\rightarrow
	R$ can be $\eta$ dependent.	Note also that the boundary region $r=R$ is	effectively a two-dimensional surface, and thus has zero measure in three-dimensions.  Thus when spatially integrating by parts in a four-dimensional integral, both $r=0$ and $r=R$ will
	be zero volume surfaces, and we can then discard any spatial surface terms. We can demonstrate this explicitly by appropriately restoring factors of $f(r)$ in the following example integral,
	\begin{equation}
		\int d^{4}x \sqrt{g}\, \mathcal{J}^{r}\nabla_{r}h=-\int d^{4}x \sqrt{g}\,h\nabla_{r}\mathcal{J}^{r}+\left(\int d^{2}\Omega d\eta\,r^{2}f(r) h \mathcal{J}^{r}\right)\bigg|^{R}_{0}.
	\end{equation}
	For non-singular functions $h$ and $\mathcal{J}^{r}$, the factor $r^{2}f(r)$ sets the spatial boundary contribution to zero.
	
	In the following calculations the function $f(r)$ arises in many places, only to be cancelled out of all final results when quantities such as $\omega_{\eta}$ above are replaced by their finite parts, eg. $\omega_{t}$. The charge flux density $J^{\eta}$ also has an apparent divergence,
	\begin{equation}
		J^{\eta}=\frac{1}{f(r)}J^{t},
	\end{equation}
however since we are assuming $J^{t}$ only has support for $r\ll R$,  we effectively have $J^{\eta}=J^{t}$.  Rather than redundantly tracking the factors of $f(r)$ through the calculation, we can simply make the $r\ll R$ approximation at the level of the metric, effectively setting $f(r)=1$, and yielding the expression
	\begin{eqnarray}
		ds^{2} &=& -d\eta^{2}+2\frac{\eta}{R}d\eta dr  +
		\bigg(1-\frac{\eta^{2}}{R^{2}}\bigg)dr^{2}+r^{2}d\Omega^{2}.
	\end{eqnarray}

\begin{figure}
		\begin{center}
			\includegraphics[width=0.4\textwidth]{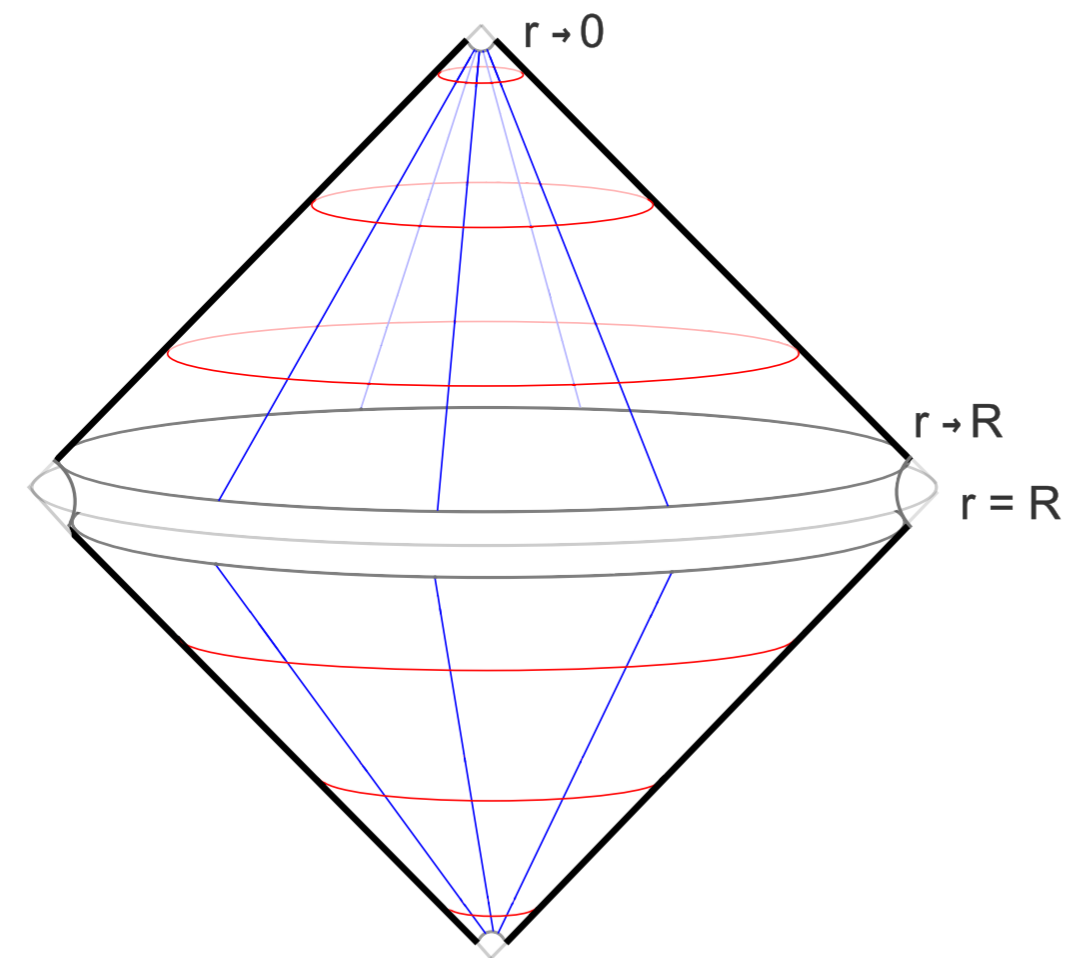}
		\end{center}
		\caption{The blow-up procedure which treats the corners of the causal
			diamond geometry. The black lines show regions of the boundary, the lightest
			grey lines show the true causal diamond, and the darker grey lines show the
			boundaries of the boundary, which in the infinite limit coincide with the true
			causal diamond. Blue lines follow null generators of the boundary, while red
			lines denote constant $t$ cuts.}\label{fig:blowup}
	\end{figure}

	This deals with the singular behaviour of the spatial ``corner'' of the
	boundary hypersurface, but there are still the corners at the top and bottom of
	the causal diamond, $r=0,\eta=\pm R$. We will also formally blow up these points
	to allow fields to take angle dependent limits as $r\rightarrow0$ on the
	boundary; see FIG. \ref{fig:blowup}. In doing this, we assume nothing enters or
	leaves $\mathcal{V}$ through the strict points $r=0,\,\eta=\pm R$.

	If one now considers a QED propagator with information specified on the
	boundary of this causal diamond, the transformation of the component $A_{\eta}$
	involves $\partial_{\eta}\Lambda$, ie., a derivative normal to surfaces of
	constant $\eta$ and thus independent of the actual pullback of $\Lambda$ to the
	surface.  Thus any boundary data specified for $A_{\eta}$ in the path integral
	will be superfluous. In addition the QED Lagrangian will be quadratic in
	$A_{\eta}$, allowing it to be integrated out via Gaussian saddle point
	substitution.

	\subsubsection{Transformed Effective Action}

	For brevity we just consider the gauge field coupled to a conserved external
	source $J^{\mu}$; this is easily generalized to scalar charged particles or to a
	Dirac field by promoting $J^{\mu}$ in the resulting boundary phase to an
	operator. As before, we first obtain results without explicitly fixing a gauge,
	then discuss how the bFP trick shortcuts the computation. Expanding the action
	so as to explicitly write $A_{\eta}$ we have
	\begin{eqnarray}
		\label{eq:causaldiamondaction}
		S &=&
		-\frac{1}{2}\int_{\mathcal{V}}d^{4}x\sqrt{g}\bigg[F^{jk}\partial_{j}A_{k}-2A_{j}J^{j}+F^{\eta
			j}\partial_{\eta}A_{j}  - \;
		A_{\eta}\bigg(\frac{1}{\sqrt{g}}\partial_{j}(\sqrt{g}F^{j\eta})+J^{\eta}\bigg)-A_{\eta}J^{\eta}\bigg]
		\qquad
	\end{eqnarray}
	where $\sqrt{g}=r^{2}\sin\theta$, and $j=\{r,\theta,\phi\}$. In writing this
	we've already freely integrated by parts in spatial directions. To integrate out
	$A_{\eta}$, we need to solve its saddle point equation, ie.
	\begin{equation}
		\frac{1}{\sqrt{g}}\partial_{j}(\sqrt{g}F^{j\eta})+J^{\eta}=0.
	\end{equation}
	
	Since the metric is non-diagonal, the resulting equation is qualitatively
	different from the previous equation for $A_{0}$. In terms of $A_{\eta}$ the
	equation of motion reads
	
	\begin{align}
		\label{eq:causaldiamondmaxwelleqn}
		\partial_{r}\big(\sqrt{g}\partial_{r}A_{\eta}\big)-g^{\eta\eta}\partial_{A}\big(\sqrt{g}g^{AB}\partial_{B}A_{\eta}\big)
		\;\;\;=\;\;\;\; &g^{\eta
			r}\partial_{A}\big(\sqrt{g}g^{AB}F_{Br}\big)+\sqrt{g}J^{\eta} \nonumber \\
		&+ \,
		\partial_{r}\big(\sqrt{g}\partial_{\eta}A_{r}\big)-g^{\eta\eta}\partial_{A}\big(\sqrt{g}g^{AB}\partial_{\eta}A_{B}\big)
	\end{align}
	
	On the right-hand side the first two terms are obviously gauge invariant,
	and the last two terms together transform as required so that the solution to
	this equation, $\tilde{A}_{\eta}$, will transform as
	$\delta_{\Lambda}\tilde{A}_{\eta}=\partial_{\eta}\Lambda$.
	
	Note that $\partial_{\eta}g^{\eta\eta}=2\eta/R^{2}$, a dimensionful quantity
	of order $R^{-1}$. By our original assumptions, $R$ is parametrically much
	larger than any other dimensionful quantity and thus this entire term is
	sub-leading. With $R$ sufficiently large we can simply assume
	$\partial_{\eta}g^{\eta\eta}=0$, allowing (\ref{eq:causaldiamondmaxwelleqn}) to
	be written compactly as
	\begin{equation}
		\label{eq:causaldiamondmaxwelleqn2}
		D^{j}\partial_{j}A_{\eta}=\frac{1}{\sqrt{g}}g^{\eta
			r}\partial_{A}\big(\sqrt{g}g^{AB}F_{Br}\big)+J^{\eta}+\partial_{\eta}D^{j}A_{j},
	\end{equation}
	where we've defined the divergence-like differential operator $D^{j}$,
	acting as
	\begin{equation}	
		D^{j}w_{j}=\frac{1}{\sqrt{g}}\partial_{r}(\sqrt{g}w_{r})-g^{\eta\eta}\frac{1}{\sqrt{g}}\partial_{A}(\sqrt{g}g^{AB}w_{B}).
	\end{equation}
	
	Now (\ref{eq:causaldiamondmaxwelleqn}) can formally be solved by assuming a
	Green's function $G$ satisfying
	\begin{equation}
		D^{j}\partial_{j}G(x,x')=\frac{\delta^{3}(x-x')}{\sqrt{g}},
	\end{equation}
	that is,
	\begin{equation}\label{eq:aetatilde}
		\tilde{A}_{\eta}=\tilde{A}^{I}_{\eta}+g+h,
	\end{equation}
	where
	\begin{equation}\label{eq:aetatildeinvar}
		\tilde{A}^{I}_{\eta}=\int_{\Sigma_{\eta}}d^{3}x'\sqrt{g}G\bigg[\frac{1}{\sqrt{g}}g^{\eta
			r}\partial_{A}\big(\sqrt{g}g^{AB}F_{Br}\big)+J^{\eta}\bigg] \qquad
	\end{equation}
	\begin{equation}\label{eq:geta}
		g=\partial_{\eta}\int_{\Sigma_{\eta}}d^{3}x'\sqrt{g} G\,D^{j}A_{j},
	\end{equation}
	and $h$ is a homogeneous solution $D^{j}\partial_{j}h=0$. Note that The
	integration in these expressions is over $\Sigma_{\eta}$, the constant $\eta$
	hypersurface corresponding to the time $\eta$ at which $\tilde{A}_{\eta}$ is
	being evaluated.
	
	We don't have a general expression for this Green's function; however the
	results that we're interested in will ultimately only depend on its value on the
	null boundary, and one can find $G$ on this boundary as well as at $\eta=0$. At
	$\eta=0$, $g^{\eta\eta}=-1$, and the differential operator simplifies to
	\begin{eqnarray}
		D^{j}\partial_{j}f(x)\big|_{\eta=0} &=&\frac{1}{\sqrt{g}} \big{[}
		\partial_{r}\big(\sqrt{g}\partial_{r}f(x)\big) + \,
		\partial_{A}\big(\sqrt{g}g^{AB}\partial_{B}f(x)\big) \big{]} \qquad
	\end{eqnarray}
	which is of course just the standard Laplacian in spherical coordinates.
	This is because the hypersurface $\eta=0$ is just the hypersurface $t=0$. Thus
	at $\eta=0$ the Green's function is given by
	\begin{equation}
		G(x,x')\big|_{\eta=0}=-\frac{1}{4\pi}\frac{1}{|x-x'|}.
	\end{equation}
	
	At the boundary, the operator $D^{j}\partial_{j}$ simplifies considerably
	because $g^{\eta\eta}$ vanishes; we then have
	\begin{equation}
		D^{j}\partial_{j}f\big|_{\eta=\pm
			R}=\frac{1}{\sqrt{g}}\partial_{r}\big(\sqrt{g}\partial_{r}f\big),
	\end{equation}
	which can be immediately integrated to find the Green's function
	\begin{equation}\label{eq:boundaryGreensfnc}
		G(x,x')\big|_{\eta=\pm
			R}=\frac{\delta^{2}(x^{A}-x^{A\,\prime})}{\sin\theta}\theta(r'-r)\bigg[\frac{1}{r}-\frac{1}{r'}\bigg].
	\end{equation}
	which propagates along the null generators of the boundary.
	
	Note that the boundary condition for $G$ is chosen so that influence
	propagates towards smaller radii, ie. causally on the future portion of
	$\partial\mathcal{V}$. When considering the past portion of
	$\partial\mathcal{V}$ one must flip the argument of the step function
	appropriately.
	
	More progress can be made when looking at the homogeneous solutions. A
	general homogeneous solution, $D^{j}\partial_{j}h=0$, will have the form
	\begin{eqnarray}
		\label{eq:causaldiamondhomogeneous}
		h(x) &=&
		\sum_{m,l}Y^{m}_{l}(\theta,\phi)\big[c^{1}_{ml}(\eta)r^{-\frac{1}{2}+\sqrt{\frac{1}{4}-g^{\eta\eta}l(l+1)}}
		+ \, c^{2}_{ml}(\eta)r^{-\frac{1}{2}-\sqrt{\frac{1}{4}-g^{\eta\eta}l(l+1)}}\big]
		\qquad
	\end{eqnarray}
	with $Y^{m}_{l}$ a spherical harmonic and $c^{1,2}_{ml}$ a set of time
	dependent coefficients.
	
	We can immediately set $c^{2}_{ml}=0$, since it is the coefficient of a term
	which will never be regular at the origin. The other term will either grow
	monotonically with $r$ or be constant in $r$. With our assumptions that the
	fields vanish at large $r$, both situations are unacceptable and we can set
	$c^{1}_{ml}$=0. The solution (\ref{eq:aetatilde}) with $h=0$ is then the unique
	solution satisfying the boundary conditions.
	
	As an aside, note that if we relax the asymptotic spatial boundary
	conditions and simply demand for the fields to be finite as
	$r\rightarrow\infty$, we can accept solutions that are independent of $r$. Such
	solutions satisfy
	\begin{equation}
		-g^{\eta\eta}l(l+1)=0.
	\end{equation}
	For all spacelike slices, $g^{\eta\eta}<0$, and the only solution is $l=0$,
	ie. a constant function of $\theta,\phi,r$. These are the time dependent global
	$U(1)$ rotations. However on the null boundaries $g^{\eta\eta}=0$, and the
	homogeneous solution space is enlarged to include any function on the sphere.
	This is interesting in the context of large gauge transformations, soft photons,
	etc, and we will return to this point in section 5.
	
	Returning to the solution (\ref{eq:aetatilde}), note that the gauge-variant
	part $g$ transforms as $\delta_{\Lambda}g=\partial_{\eta}\Lambda$. From
	(\ref{eq:geta}) we see we can identify it as a $g$-potential of form
	$g=\partial_{\eta}\Phi$ with $\Phi$ given by
	\begin{equation}\label{eq:etagpotential}
		\Phi(x)=\int_{\Sigma_{\eta}}d^{3}x'\sqrt{g}\,G(x,x')\,D^{j}A_{j}(x').
	\end{equation}
	
	For the causal diamond we can now decompose the gauge field into a
	gauge-invariant part $\mathcal{A}_{j}=A_{j}-\partial_{j}\Phi$, and a pure gauge
	part $\partial_{j}\Phi$; the subsequent development then parallels to that for
	the time slice.  We substitute $\tilde{\mathcal{A}}_{\eta}$ into the action
	(\ref{eq:causaldiamondaction}) and rewrite the action in the new variables
	$\mathcal{A}_{j},\Phi$. Using current conservation, we then get an effective
	action
	
	\begin{align}
		\tilde{S}\;\;=\;\; \int_{\partial\mathcal{V}}d^{3}x\sqrt{g}\Phi
		J^{\eta}& -\tfrac{1}{2}\int_{\mathcal{V}}d^{4}x\sqrt{g}\bigg[\tilde{F}^{\mu
			j}\partial_{\mu} \mathcal{A}_{j}-2\mathcal{A}_{j}J^{j} \nonumber \\	
		&-J^{\eta}\int_{\Sigma_{\eta}}d^{3}x'\sqrt{g}G\bigg(J^{\eta}+\frac{1}{\sqrt{g}}g^{\eta
			r}\partial_{A}(\sqrt{g}g^{AB}F_{Br})\bigg)\bigg],
	\end{align}
	with
	\begin{equation}
		\tilde{F}^{\mu
			j}=\partial^{\mu}\tilde{\mathcal{A}}^{j}-\partial^{j}\tilde{\mathcal{A}}^{\mu}.
	\end{equation}
	
	Note that all of the terms involving $\Phi$ again summed to a total time
	derivative, and thus formed a boundary term in the action. The remaining bulk
	action is written in terms of explicitly gauge invariant variables.
	
	We can actually take this expression further because the variable
	$\mathcal{A}_{j}=A_{j}-\partial_{j}\Phi$ is actually transverse in the sense
	that $D^{j}\mathcal{A}_{j}=0$. Using this, and a few spatial integrations by
	parts, we expand the effective action in terms of the gauge invariant variables
	to get
	\begin{align}
		\label{eq:causaldiamondaction2}
		\tilde{S}=&\int_{\partial\mathcal{V}}d^{3}x\sqrt{g}\Phi J^{\eta} \;+\;
		\tfrac{1}{2}\int_{\mathcal{V}}d^{4}x\sqrt{g}\bigg[\partial_{\eta}\mathcal{A}_{r}\partial_{\eta}\mathcal{A}_{r}-g^{\eta\eta}g^{AB}\partial_{\eta}\mathcal{A}_{A}\partial_{\eta}\mathcal{A}_{B}-2g^{\eta
			r}g^{AB}(\partial_{\eta}\mathcal{A}_{A})F_{rB} \nonumber \\
		& \qquad\qquad
		-F^{AB}\partial_{A}A_{B}+g^{AB}F_{rA}F_{rB}+2\mathcal{A}_{j}J^{j} \nonumber \\ &
		\qquad\qquad + \bigg(J^{\eta}+\frac{1}{\sqrt{g}}g^{\eta
			r}\partial_{A}(\sqrt{g}g^{AB}F_{Br})\bigg)\int_{\Sigma_{\eta}}d^{3}x'\sqrt{g}G\bigg(J^{\eta}+\frac{1}{\sqrt{g}}g^{\eta
			r}\partial_{C}(\sqrt{g}g^{CD}F_{Dr})\bigg)\bigg]
	\end{align}

	This is the expression we will work with - although we will not actually
	perform computations with this action. The purpose of the derivation was rather
	to demonstrate that when propagators are considered for different boundary
	geometries, we can still unambiguously extract a boundary term describing the
	dressing required to make charged states gauge invariant.
	
	As expected the action (\ref{eq:causaldiamondaction2}) is non-local in
	space. The ``Coulomb'' interaction term now contains not just the charge density
	$J^{\eta}$ but also terms describing the magnetic field. These apparent
	interactions arise because our coordinates are no longer adapted to the
	isometries of Minkowski spacetime. One can set $\eta=0$, and thus $g^{\eta
		r}=0$, to verify that on a standard constant time slice, this gives the usual
	Lagrangian  density.

	\subsection{Form of the Propagator}
	\label{sec:diamond-K}
	
	
	Since $\Phi$ doesn't appear in the integrand, the path integral over $\Phi$
	can again be removed by a Faddeev-Popov procedure. The propagator
	\begin{equation}
		K(A_{\mu\,\partial\mathcal{V}})=\int_{A_{\mu\,\partial\mathcal{V}}}\mathcal{D}A_{\mu}\,e^{i\int_{\mathcal{V}}d^{4}x\sqrt{g}\big[-\frac{1}{4}F_{\mu\nu}F^{\mu\nu}+A_{\mu}J^{\mu}\big]}
	\end{equation}
	for evolution of the gauge field coupled to a source $J^{\mu}$, through a
	large causal diamond, is then equal to
	\begin{equation}
		K(A_{\mu\,\partial_{V}})=e^{i\int_{\partial\mathcal{V}}J^{\eta}\Phi}\int_{\mathcal{A}_{j\,\partial\mathcal{V}}}\mathcal{D}
		\mathcal{A}_{j}\,e^{i\tilde{S}[\tilde{\mathcal{A}}|J]},
	\end{equation}
	where the effective action $\tilde{S}$ is given by the bulk part of
	(\ref{eq:causaldiamondaction2}), and the prefactor involves the generalized
	Coulomb dressing, in which $\Phi$ is given by eqns. (\ref{eq:etagpotential}) and
	(\ref{eq:boundaryGreensfnc}) evaluated on the boundary. The contribution from
	the future part reads
	\begin{equation}\label{eq:gpotentialnull}
		\Phi\big|_{\partial\mathcal{V}}(r',x^{\prime
			A})=\int_{r^{\prime}}^{\infty} dr
		\bigg(\frac{1}{r^{\prime}}-\frac{1}{r}\bigg)
		\partial_{r}\big(r^{2}A_{r}(r,x^{\prime A})\big),
	\end{equation}
	whereas on the past part the integration is over all $r$ interior to
	$r^{\prime}$.
	
	This dressing describes the radial electric field at each point on
	$\partial\mathcal{V}$, with a strength determined by the total charge flux
	through $\partial\mathcal{V}$ at earlier times. This is our central result for
	the causal diamond geometry.
	
	We emphasize again that this result is not the result of a specific gauge
	choice, and that the definition of gauge-invariant variables $\mathcal{A}_{j}$
	again emerged naturally from the path integral. Remarkably, our procedure
	succeeded even though there is no unambiguous notion of the `transverse' vector
	field, since one cannot define an intrinsic divergence on a null boundary.
	
	If we now give the matter current $J^{\mu}$ its own dynamics, we can easily
	generalize the above derivation. This is possible because $U(1)$ charge current
	is conserved off shell for particles. Alternatively, as before,  we can go back
	and skip the step which invokes current conservation by using the bFP trick. The
	derivations are as before; for Dirac fermions we then get the gauge invariant
	QED amplitude on the large causal diamond to be
	
	\begin{align}
		\label{eq:QEDcausaldiamondampl}
		K(A_{\mu\,\partial_{V}},\psi_{\partial\mathcal{V}})\;\;=\;\;\;
		\int_{e^{-ie\Phi}\psi_{\partial\mathcal{V}}}\mathcal{D}\psi\mathcal{D}\bar{\psi}\int_{\mathcal{A}_{j\,\partial\mathcal{V}}}\mathcal{D}\mathcal{A}_{j}\,e^{iS[\mathcal{A}|J]-i\int_{\mathcal{V}}d^{4}x\sqrt{g}\,\bar{\psi}\big(\gamma^{\mu}\partial_{\mu}+m\big)\psi}
	\end{align}
	
	where $\Phi$ is given by (\ref{eq:gpotentialnull}). Analogous to the
	time-slice amplitude we see a dressing of each Dirac excitation in the boundary
	state by a Coulombic electric field.
	
	Since we have skipped the explicit derivation of
	(\ref{eq:QEDcausaldiamondampl}) and foregone the discussion of general
	boundaries in curved spacetime, we should at least mention that to do the bFP
	trick for more general boundaries one must necessarily use generalizations of
	canonical conjugate momenta and commutation relations. To highlight this, for a
	general path integral with data specified on boundary $\partial\mathcal{V}$, we
	can consider a variation of this boundary data, viz.,
	\begin{equation}
		\delta\int_{\phi_{\partial\mathcal{V}}}\mathcal{D}\phi\,e^{iS[\phi]}=i\int_{\phi_{\partial\mathcal{V}}}\mathcal{D}\phi\,e^{iS[\phi]}\delta
		S.
	\end{equation}
	A general variation of the action is of the form
	\begin{equation}\label{eq:varyaction}
		\delta
		S=\int_\mathcal{V}d^{4}xE(\phi)\delta\phi+\int_{\partial\mathcal{V}}d^{3}x\,(\partial_{\mu}S)\theta^{\mu}(\phi,\delta\phi),
	\end{equation}
	where $E(\phi)$ is the scalar density equation of motion, the boundary is
	defined by a constant $S$ hypersurface, and the symplectic potential current
	density $\theta^{\mu}$ is given for a general Lagrangian in ref.
	\cite{lee-wald}. For a Lagrangian density which is a function only of the fields
	and their first derivatives we have
	\begin{equation}		
		\theta^{\mu}(\phi,\delta\phi)=\frac{\partial\mathcal{L}}{\partial\nabla_{\mu}\phi}\delta\phi.
	\end{equation}
	For non-null boundaries  $\sqrt{g}\partial_{\mu}S$ can be related to the
	normal covector and intrinsic volume element for the hypersurface, but the form
	in (\ref{eq:varyaction}) is more general and also applies to null boundaries.
	
	For variations with support only on the boundary we then have the functional
	derivative
	
	\begin{equation}
		\label{eq:symplecticcurrentdensity}
		\frac{\delta}{\delta\phi_{\partial\mathcal{V}}(x)}
		\int_{\phi_{\partial\mathcal{V}}}\mathcal{D}\phi\,e^{iS[\phi]}  \;\;\; =\;\;\;
		i\int_{\phi_{\partial\mathcal{V}}}\mathcal{D}\phi\,e^{iS[\phi]}\bigg[\int_{\partial\mathcal{V}}d^{3}x'\,\frac{\delta\theta^{S}(\phi,\delta\phi)}{\delta\phi(x)}\bigg]
	\end{equation}

	Defining $\frac{\delta}{\delta\phi(x)}\phi(x')=\delta^{3}(x-x')/\sqrt{g}$,
	the commutation relation between $\phi$ and $-i\delta/\delta\phi$ is obviously
	canonical. The functional derivatives
	$\frac{-i\delta}{\delta\phi_{\partial\mathcal{V}}}$ used in the bFP trick will
	then be operator representations of the generalized conjugate momentum
	\begin{equation}		
		\Pi_{\partial\mathcal{V}}(x)=\int_{\partial\mathcal{V}}d^{3}x'\,\theta^{S}(\phi,g^{-1/2}\delta^{3}(x-x')).
	\end{equation}
	This expression was used in deriving (\ref{eq:QEDcausaldiamondampl}), and
	will be explicitly used in the following section.

	
	\section{Large Gauge Transformations and Additional Constraints}
	\label{sec:largegauge}
	
	
	Up to now we have assumed that both $A_{\mu}$ and the gauge transformations
	on $A_{\mu}$ vanish sufficiently fast at spatial infinity that one can freely
	integrate by parts any expression with spatial derivatives. Energy-flux
	finiteness arguments lead one  to expect the field strength $F_{\mu\nu}$ to obey
	such asymptotic fall-off conditions, at least in many physical situations.
	However, it is not clear why either $A_{\mu}$, or gauge transformations of
	$A_{\mu}$, should vanish at infinity.
	
	Gauge transformations which don't fall off as quickly as required for the
	above manipulations are referred to as large gauge transformations. These have a
	long history, especially in gravity \cite{BondiS}, and have been extensively
	discussed in recent years \cite{strominger,largeGT,IRstuff}. Many different
	choices of asymptotic fall-off conditions for $A_{\mu}$  have been made in the
	literature.
	
	Invariance under the set of large gauge transformations implies a further
	set of constraints, in addition to Gauss' law and
	$\hat{E}^{0}\Psi=i\tfrac{\delta}{\delta A_{0}}\Psi=0$. In this section we
	enlarge the set of allowed gauge transformations to those which are finite and
	non-vanishing at the spatial boundary, and generalize the techniques used above
	to handle these. The invariant propagators then shed light on the constraints
	implied by large gauge invariance; and the path integral gives explicit
	solutions to the operator constraint equations.
	
	We will treat the spatial boundary as a large sphere or cylinder of radius
	$R\rightarrow\infty$, and we allow for gauge transformations which have finite
	asymptotic limits, viz.,
	\begin{equation}
		\lambda(t,x^{A})\equiv\lim_{r\rightarrow R} \Lambda(t,r,x^{A}).
	\end{equation}
	
	With finite asymptotic limits for $\Lambda$, we must also allow for finite
	asymptotic limits for the gauge field, viz.,
	\begin{equation}
		a_{\mu}(t,x^{A})\equiv\lim_{r\rightarrow R} A_{\mu}(t,r,x^{A}).
	\end{equation}
	
	In what follows, we warm up by discussing, in section 4.1, large gauge
	transformations for propagation between time slices; we then proceed in section
	4.2 to the causal diamond.

	\subsection{Large Gauge Transformations: Time Slicing}
	\label{sec:largeG-TS}

	We would like to compute the propagator
	\begin{equation}
		\label{eq:largetime-sliceprop}		
		K(A_{\mu\,\partial\mathcal{V}})=\int_{{A_{\mu\,\partial\mathcal{V}}}}\mathcal{D}A_{\mu}\,e^{iS},
	\end{equation}
	where the region $\mathcal{V}$ over which we integrate is again part of
	Minkowski space, bounded by the constant $t$ slices $\Sigma_{i}$, $\Sigma_{f}$
	and the large cylinder of radius $R\rightarrow\infty$, $\Sigma_{\infty}$, and
	the action is just (\ref{eq:time-sliceaction}). Again, for brevity we assume
	that the source is an external conserved current, but as was the case in the
	first section, the following manipulations easily generalize to dynamic matter
	fields. As just discussed, while we fix boundary data
	$A_{\mu\,\partial\mathcal{V}}$ on all of $\partial\mathcal{V}$, we now lift the
	restriction that $A_{\mu}$ vanishes at spatial infinity.
	
	At the technical level, the new challenge is that we can no longer uniquely
	invert the Laplacian operator when solving the Gauss law equation as in eqn.
	(\ref{eq:A0tilde}); there is now nothing restricting the homogeneous solutions.
	
	To proceed with the integral we need to again use the boundary Faddeev-Popov
	trick, in the form (\ref{eq:FPintegral}). Suppose now that one tries to fix a
	Coulomb gauge in the FP path-integral, ie., write
	$\mathcal{G}(A)=\partial^{j}A_{j}$. However in the enlarged gauge group this
	choice will leave the gauge under-determined, because there are homogeneous
	solutions, $\nabla^{2}\Lambda=0$ which are non-vanishing at spatial infinity.
	
	If however we restrict ourselves to gauge functions which are finite at
	spatial infinity, then the only remaining homogeneous solution is
	$\Lambda(x)=c(t)$. The only residual gauge transformations in the FP integral
	(\ref{eq:FPintegral}) are then time dependent global $U(1)$ rotations. These
	leave the spatial components $A_{j}$ invariant, and only shift the spatially
	constant part of $A_{0}$. To properly implement the bFP trick we then must
	supplement the Coulomb gauge fixing delta function with another delta function
	which eliminates these residual transformations.
	
	A sufficient choice is to gauge fix the $l=0$ spherical harmonic mode of the
	asymptotic gauge function $\lambda(t,x^{A})$. We refer to the $l=0$ part of a
	function on the sphere using a superscript ``$(0)$''. Up to field-independent
	normalization we may then write
	\begin{equation}\label{eq:largetime-sliceFP}		
		1=\int\mathcal{D}\Lambda\,\delta(\partial^{j}A_{j}^{\Lambda})\delta(a^{(0)\,\Lambda}_{0}).
	\end{equation}
	in place of (\ref{eq:FPintegral}).
	
	In what follows it is more clear if we explicitly separate out the
	asymptotic $l=0$ part of all functions. The notation may seem heavier than
	necessary but it will allow for a much quicker generalization to the later
	treatment of the causal diamond amplitude. We will therefore write,
	\begin{equation}
		\Lambda(t,r,x^{A})=\bar{\Lambda}(t,r,x^{A})+\lambda^{(0)}(t),
	\end{equation}
	where $\bar{\Lambda}$ has a finite asymptotic limit
	$\bar{\lambda}(t,x^{A})=\lim_{r\rightarrow\infty}\bar{\Lambda}(t,r,x^{A})$, but
	the function $\bar{\lambda}(t,x^{A})$ has a vanishing $l=0$ mode. We'll use this
	same notation for the gauge field, in terms of which the action is simply
	\begin{equation}
		S[A]=S[A_{j},\bar{A}_{0}]+\int_{t_{i}}^{t_{f}}dt a^{(0)}_{0}Q,
	\end{equation}
	where $Q=\int d^{3}xJ^{0}$ is the total charge.
	
	With this, we can now multiply the propagator (\ref{eq:largetime-sliceprop})
	by a carefully chosen factor of 1, from (\ref{eq:largetime-sliceFP}), to obtain
	
	\begin{eqnarray}
		K(A_{\partial\mathcal{V}})  \;\;=\;\;  \int\mathcal{D}\bar{\Lambda} \,
		d\lambda^{(0)}\int_{A_{\mu\,\partial\mathcal{V}}}\mathcal{D}\bar{A}_{0}
		\,\mathcal{D}a^{(0)}_{0} \,\mathcal{D}A_{j}
		\; \delta(\partial^{j}A_{j}^{\Lambda})
		\,\delta(a^{(0)\,\Lambda}_{0})\;e^{iS[A_{j},\bar{A}_{0}]+i\int_{t_{i}}^{t_{f}}
			dt\, a_{0}^{(0)}Q}.
	\end{eqnarray}
	
	Now, we implement the bFP trick by changing variables, as done before (cf.
	eqns.  (\ref{eq:bFPstep1})-(\ref{eq:bFPstep3}), and (\ref{eq:bFPstep4})), to get
	\begin{align}
		\label{eq:time-slicelargebFP}
		K(A_{\partial\mathcal{V}})&\;\;=\;\;
		\int\mathcal{D}\bar{\Lambda}d\lambda^{(0)}\,\delta^{\partial\mathcal{V}}(\partial^{j}A_{j}+\nabla^{2}\bar{\Lambda})\delta^{\partial\mathcal{V}}(a^{(0)}_{0}+\partial_{0}\lambda^{(0)})
		\nonumber \\
		& \qquad\qquad\qquad  \times \;
		e^{-i\int_{\partial\mathcal{V}}\big[\bar{\Lambda}J^{0}+\lambda^{(0)}J^{0}+i\partial_{0}\bar{\Lambda}\frac{\delta}{\delta
				\bar{A}_{0}}+i\partial_{0}\lambda^{(0)}\frac{\delta}{\delta
				a^{(0)}_{0}}+i\partial_{j}\bar{\Lambda}\frac{\delta}{\delta A_{j}} \big]}
		\nonumber \\
		& \qquad\qquad\qquad\qquad\qquad \times \;
		\int_{A_{\mu\,\partial\mathcal{V}}}\mathcal{D}\bar{A}_{0}\mathcal{D}a^{(0)}_{0}\mathcal{D}A_{j}\,\delta(\partial^{j}A_{j})\delta(a^{(0)}_{0})\,e^{iS[A_{j},\bar{A}_{0}]+i\int_{t_{i}}^{t_{f}}
			dt\, a_{0}^{(0)}Q}
	\end{align}
	
	In the bulk part of the path integral we have effectively inserted gauge
	fixing delta functions as desired. The additional gauge fixing delta function
	simply sets $a^{(0)}_{0}=0$, reducing the action to its usual form. As always in
	the bFP trick, we've also obtained a number of delta functions and linear shift
	operators outside the path integral. The crucial observation here is that the
	delta functions constraining the boundary gauge transformations constrain only
	$\bar{\Lambda}$ and $\partial_{0}\lambda^{(0)}$, they do not constrain the other
	independent functions $\partial_{0}\bar{\Lambda}$ and $\lambda^{0}$.
	
	In factoring out the bulk gauge group integral we are then left with
	residual integrals over $\partial_{0}\bar{\Lambda}$ and $\lambda^{0}$. The
	remaining boundary integrals over $\bar{\Lambda}$ and
	$\partial_{0}\lambda^{(0)}$ are trivially performed using the delta functions.
	The result is then
	\begin{equation}
		K(A_{\partial\mathcal{V}})=\bigg(\int
		d\lambda^{(0)}e^{-i\int_{\partial\mathcal{V}}\lambda^{(0)}J^{0}}\bigg)e^{i\int_{\partial\mathcal{V}}\nabla^{-2}(\partial^{j}A_{j})J^{0}}
		\int_{\tilde{\mathcal{A}}_{j\,\partial\mathcal{V}}}\mathcal{D}\bar{A}_{0}\mathcal{D}A_{j}\,\,\delta(\partial^{j}A_{j})e^{iS[A_{j},\bar{A}_{0}]},
	\end{equation}
	
	where $\mathcal{A}_{j}$ is the transverse component of $A_{j}$. We can now
	perform the $\bar{A}_{0}$ integral and there is no ambiguity in its saddle point
	solution; it is again given by $\tilde{A}_{0}=\nabla^{-2}J^{0}$ and the
	homogeneous solution is necessarily zero because by definition $\bar{A}_{0}$ has
	vanishing asymptotic $l=0$ mode.
	
	Note the remarkable feature, that the vestige of working on the
	configuration space for $A^{\mu}$ with non-vanishing asymptotic limit is just
	the integral over $\lambda^{(0)}$ on the boundary. Since $\lambda^{(0)}$ was
	time-dependent but spatially constant function, these are independent integrals
	on each of the past and future timeslices of the charge density.  This does
	nothing other than add a delta function enforcing charge neutrality on the
	boundary state. In hindsight it is completely obvious that if we demand the
	amplitude to be invariant under time-dependent global $U(1)$ transformations,
	the state must be charge neutral - by enlarging the gauge group, we've simply
	imposed this new constraint.

	In principle one could never determine whether the total charge
	on a spatial slice is zero, since for any region with non-zero charge there
	could be compensating charges arbitrarily far away, which would ensure the
	overall charge neutrality condition.  This is always possible in $U(1)$ gauge
	theory, because the interaction becomes arbitrarily weak at large distance.
	
	Rather than limiting ourselves to such a scenario with fictitious image
	charges as infinity, we could instead restrict the path-integral such that we
	consider only gauge fields which are time-independent on the spatial boundary.
	Classically this corresponds to enforcing the electric field to vanish as
	$r\rightarrow\infty$, and is thus a sensible restriction if one wants to
	consider quantum fluctuations about finite energy classical background
	configurations. In this case, the allowed gauge transformations have asymptotic
	limits which are simply time-independent functions on a sphere
	\begin{equation}
		\lambda(x^{A})\equiv\lim_{r\rightarrow R}\Lambda(t,r,x^{A}).
	\end{equation}
	
	Even with these time-independent asymptotic limits, there is still a
	residual gauge-fixing to be done in addition to specifying the Coulomb gauge.
	The difference now, compared with eq. (\ref{eq:largetime-sliceFP}), is that we
	only need to gauge fix the $l=0$ spherical harmonic mode of the asymptotic gauge
	field on a single time-slice. The analogous calculations above are identical in
	form, so we will simply quote the final result below.
	
	If gauge fields are allowed to reach finite, but constant-in-time,
	configurations at spatial infinity, then the only allowed large gauge
	transformation in the path-integral is just a single global $U(1)$ phase
	rotation. The resulting propagator is then
	\begin{equation}
		K(A_{\partial\mathcal{V}})=\bigg(\int
		d\lambda^{(0)}e^{-i\lambda^{(0)}\Delta
			Q}\bigg)e^{i\int_{\partial\mathcal{V}}\nabla^{-2}(\partial^{j}A_{j})J^{0}}
		\int_{\tilde{\mathcal{A}}_{j\,\partial\mathcal{V}}}\mathcal{D}\bar{A}_{0}\mathcal{D}A_{j}\,\,\delta(\partial^{j}A_{j})e^{iS[A_{j},\bar{A}_{0}]},
	\end{equation}
	where
	\begin{equation}
		\Delta Q =
		\int_{\partial\mathcal{V}}J^{0}=\int_{\Sigma_{f}}d^{3}x\,J^{0}-\int_{\Sigma_{i}}d^{3}x\,J^{0}.
	\end{equation}
	This result is identical to the case eq. (\ref{eq:finalprop}) where large
	gauge transformations were forbidden, except here total charge conservation
	is being explicitly enforced.
	
	This can be summarized succinctly. Previously we had required invariance
	under global $U(1)\times U(1)$ transformations of our boundary data, where
	each the $U(1)$ acts independently on the past and future surfaces
	($\Sigma_{i}, \Sigma_{f}$), and this then required the total charge on each
	of $\Sigma_{i}$ and $\Sigma_{f}$ to vanish. Upon restricting the allowed
	gauge field configurations to be time-independent on the spatial boundary
	$\Sigma_{B}$, we then only required invariance under the diagonal subgroup
	of global $U(1)$ rotations which act simultaneously on both $\Sigma_{i}$ and
	$\Sigma_{f}$, and consequently it was the difference between total charges
	on $\Sigma_{f}$ and $\Sigma_{i}$ which needed to vanish.

	Note that in both scenarios considered above we did not need to eliminate
	all gauge functions which are finite asymptotically in order to fix the
	residual gauge freedom; it was only necessary to gauge fix that part of the
	gauge field which was constant on the sphere at spatial infinity. Gauge
	functions which approach $l\neq0$ functions on the asymptotic sphere are
	still allowed; they simply do not affect time slice amplitudes. In the next
	subsection we see that allowing such $l\neq0$ gauge transformations
	actually has a nontrivial effect on the causal diamond amplitude.

	\subsection{Large Gauge Transformations: Causal Diamond Evolution}
	
	
	We would now like to consider the amplitude
	\begin{equation}\label{eq:amp4.2}		
		K(A_{\mu\,\partial\mathcal{V}})=\int_{{A_{\mu\,\partial\mathcal{V}}}}\mathcal{D}A_{\mu}\,e^{iS},
	\end{equation}
	where, as before, $S$ is given by the sourced Maxwell action and the
	integration region $\mathcal{V}$ is the causal diamond of radius
	$R\rightarrow\infty$, but now we allow the gauge fields to be finite as
	$r\rightarrow R$. The story is very similar to the treatment of large fields in
	the time slice propagator, but with an interesting additional feature.
	
	Looking back to (\ref{eq:causaldiamondmaxwelleqn2}) and its solution
	(\ref{eq:aetatilde}), we can see that without the assumption that the gauge
	field vanishes as $r\rightarrow\infty$, there are infinitely many possible
	homogeneous solutions. In the bulk, $\eta\in(-R,R)$, the only acceptable
	homogeneous solution (\ref{eq:causaldiamondhomogeneous}) is a time-varying
	$h(\eta)$ which is constant in space.
	
	\subsubsection{Applying the bFP trick}
	
	The situation for these time-dependent global $U(1)$ transformations is the
	same as considered above for $A_{0}$ in the time slice amplitude, and the same
	remedy applies. We must separate off the asymptotic $l=0$ part,
	$a_{\eta}^{(0)}$, and enforce an additional gauge fixing which sets
	$a_{\eta}^{(0)}=0$ in the bulk. This will allow for a unique saddle point
	solution for the remaining field $\bar{A}_{\eta}=A_{\eta}-a_{\eta}^{(0)}$. The
	upshot is the same as the previous case; properly treating this asymptotic $l=0$
	part will just introduce delta functions on the boundary which enforce overall
	charge neutrality $\int_{\partial\mathcal{V}}d^{3}x\sqrt{g}J^{\eta}=0$.
	
	However there is another, more interesting, result in the causal diamond
	geometry. When we implement the bFP trick, we aim to introduce the Faddeev-Popov
	gauge fixing as in (\ref{eq:largetime-sliceFP}) above; but the integrand here
	still does not uniquely fix the gauge. This is because when $\eta=\pm R$, ie. on
	the boundary of the causal diamond, there are homogeneous solutions
	$D^{j}\partial_{j}\Lambda=0$ which are arbitrary functions on the sphere. The
	above delta functions will uniquely determine the gauge function $\Lambda$ in
	the bulk, but on each of the future and past portions of the boundary there is
	still a residual gauge freedom given by all functions $\Lambda$ approaching a
	non-constant $(l\neq0)$ function on the sphere, $\lambda^{(l\neq0)}(x^{A})$, as
	$\eta\rightarrow\pm R$.
	
	Such functions will be discontinuous at $r=0,\eta=\pm R$, but this is
	allowed since these singular points have been formally ``blown up'', allowing
	for such angle dependent limits as $r\rightarrow0$ on the boundary.
	
	To uniquely fix the gauge we then append a further gauge fixing term on the
	boundary; we choose $\nabla^{B}a_{B}^{\Lambda}=0$, where $\nabla^{B}a_{B}$ is
	the vector divergence on the unit two-sphere. These choices together uniquely
	fix the gauge, so that we can write
	\begin{equation}\label{eq:largeGFP}
		1=\int
		\mathcal{D}\Lambda\,\delta(D^{j}A_{j}^{\Lambda})\delta(a_{\eta}^{(0)\,\Lambda})\delta^{\partial\mathcal{V}}(\nabla^{B}a_{B}^{\Lambda}),
	\end{equation}
	up to a field independent constant. To reduce the clutter in this and
	following equations, we have used the condensed notation
	$\delta^{\partial\mathcal{V}}(\nabla^{B}a_{B}^{\Lambda})=\delta^{\partial\mathcal{V}^{+}}(\nabla^{B}(a^{+}_{B})^{\Lambda})\delta^{\partial\mathcal{V}^{-}}(\nabla^{B}(a^{-}_{B})^{\Lambda})$,
	where we have introduced the notation $\partial\mathcal{V}^{\pm}$ to refer to
	the future/past portions of the causal diamond boundary respectively. We remind
	the reader that
	\begin{equation}
		a_{B}^{\pm}(x^{A})=\lim_{r\rightarrow R}A_{B}(\eta=\pm R,r,x^{A})
	\end{equation}
	is the value of the gauge field taken on the boundaries of the future/past
	boundaries respectively, denoted collectively by
	$\partial\partial\mathcal{V}$ below.
	
	Note that $\partial\mathcal{V}$ is not connected because of the blow-up
	procedure, so $\partial\partial\mathcal{V}$ is non-empty. Rather, these
	regions are approached in the $r\rightarrow \infty$ limit of the future null
	surface and the past null surface. These codimension-2 surfaces are
	effectively the past boundary of future null infinity $\mathcal{I}^{+}_{-}$
	and future boundary of past null infinity $\mathcal{I}^{-}_{+}$ commonly
	discussed in the literature of this topic~(see \cite{strominger} and refs.
	therein).
	
	Before proceeding, let us also emphasize that because the future and past
	portions of the boundary are not connected, it was necessary to fix this
	residual gauge freedom on each of the future and past surfaces
	separately \footnote{However, see footnote 2 and the discussion
		at the end of Sec. (4.2).}.
	
	We can now multiply eq. (\ref{eq:largeGFP}) into our path integral
	representation for the causal diamond propagator. Doing this and implementing
	again the bFP trick, we obtain
	
	\begin{align}
		\label{eq:causaldiamondlargebFP}
		K(A_{\partial\mathcal{V}})&
		\;\;=\;\;\int\mathcal{D}\bar{\Lambda}d\lambda^{(0)}\,\delta^{\partial\mathcal{V}}(D^{j}A_{j}+D^{j}\partial_{j}\bar{\Lambda})\delta^{\partial\mathcal{V}}(a^{(0)}_{\eta}+\partial_{\eta}\lambda^{(0)})\delta^{\partial\mathcal{V}}(\nabla^{B}a_{B}+\nabla^{B}\nabla_{B}\lambda)
		\nonumber \\
		& \qquad\qquad\qquad \times \;
		e^{-i\int_{\partial\mathcal{V}}\big[\bar{\Lambda}J^{\eta}+\lambda^{(0)}J^{\eta}+i\partial_{\eta}\bar{\Lambda}\frac{\delta}{\delta
				\bar{A}_{\eta}}+i\partial_{\eta}\lambda^{(0)}\frac{\delta}{\delta
				a^{(0)}_{\eta}}+i\partial_{j}\bar{\Lambda}\frac{\delta}{\delta A_{j}} \big]}
		\nonumber \\
		& \qquad\qquad\qquad\qquad \times
		\int_{A_{\mu\,\partial\mathcal{V}}}\mathcal{D}\bar{A}_{\eta}\mathcal{D}a^{(0)}_{\eta}\mathcal{D}A_{j}\,\delta(D^{j}A_{j})\delta(a^{(0)}_{\eta})\,e^{iS[A_{j},\bar{A}_{\eta}]+i\int_{-R}^{R}
			d\eta\, a_{\eta}^{(0)}Q}
	\end{align}
	
	The propagator (\ref{eq:causaldiamondlargebFP}) is structurally very similar
	to (\ref{eq:time-slicelargebFP}), except for the new factor we've introduced to
	gauge fix the residual transformations which are allowed on the causal diamond
	boundary. Again, we can freely evaluate the integral over
	$\partial_{\eta}\bar{\Lambda}$ and $\lambda^{(0)}$ in the boundary gauge
	transformations since they are not fixed by the boundary gauge fixing delta
	functions. We can evaluate the integrals over $\bar{\Lambda}$ and
	$\partial_{\eta}\lambda^{(0)}$ using the delta functions. The boundary delta
	functions now constrain $\bar{\Lambda}$ to be
	\begin{equation}
		\bar{\Lambda} \;\;=\;\;
		-\int_{\partial\mathcal{V}}d^{3}x\sqrt{g}GD^{j}A_{j}-\int_{\partial\partial\mathcal{V}}
		d^{2}\Omega \bar{G}\nabla^{B}a_{B}
	\end{equation}
	where, up to a minus sign, the first term is just $\Phi$ given in
	(\ref{eq:gpotentialnull}); $d^{2}\Omega$ is the area element on the unit
	two-sphere, and $\bar{G}$ is the Green's function for the Laplacian (less the
	$l=0$ mode) on the unit two-sphere. Evaluating these integrals we obtain the
	final expression for the large gauge transformation invariant causal diamond
	amplitude
	\begin{align}\label{eq:largediamondampl}
		K(A_{\partial\mathcal{V}})& \;\;=\;\; \bigg(\int
		d\lambda^{(0)}e^{-i\int_{\partial\mathcal{V}}\lambda^{(0)}J^{\eta}}\bigg)
		e^{i\int_{\partial\mathcal{V}}d^{3}x\sqrt{g}\big[\int_{\partial\mathcal{V}}d^{3}x'\sqrt{g}GD^{j}A_{j}+\int_{\partial\partial\mathcal{V}}
			d^{2}\Omega' \bar{G}\nabla^{B}a_{B})\big]J^{\eta}} \nonumber \\
		& \qquad\qquad\qquad \times
		\int_{\mathcal{A}_{j\,\partial\mathcal{V}}}\mathcal{D}\bar{A}_{\eta}\mathcal{D}A_{j}\,\delta(D^{j}A_{j})\,e^{iS[A_{j},\bar{A}_{\eta}]}
	\end{align}

	Just as we found for the time-slice geometry, there is a $\lambda^{(0)}$
	integral which enforces total charge flux neutrality,
	$Q_{\partial\mathcal{V}^{+}}=Q_{\partial\mathcal{V}^{-}}=0$,
	\begin{equation}
		Q_{\partial\mathcal{V}^{\pm}}=\int_{\partial\mathcal{V}^{\pm}}J^{\eta},
	\end{equation}
	where $\partial\mathcal{V}^{\pm}$ denotes the future/past portion of the null
	boundary. As discussed at the end of Sec. (\ref{sec:largeG-TS}), we could relax
	this condition down to total charge conservation
	$Q_{\partial\mathcal{V}^{+}}=Q_{\partial\mathcal{V}^{-}}\neq0$ by requiring the
	$l=0$ part of the gauge fields to be time-independent at spatial infinity. As
	discussed above, this global $U(1)$ invariance is not necessarily interesting.
	
	What is novel about the causal diamond amplitude eq.
	(\ref{eq:largediamondampl}) is that in addition to the total charge flux
	constraint, we've found that invariance under large gauge transformations with
	higher spherical harmonics enforces a new constraint on the system. Since
	$\bar{G}$ and $a_{B}$ are just angular functions, the new boundary phase in
	(\ref{eq:largediamondampl}) implies that a certain part of the electric field at
	each angle is determined solely by the net flux of charge through the boundary
	at each angle. To see this explicitly,  we define
	\begin{equation}
		\frac{\delta}{\delta
			a^{\sigma}_{A}(x^{A})}a^{\sigma'}_{B}(x^{A\prime})=\delta^{\sigma'}_{\sigma}\delta^{A}_{B}\,q^{-1/2}\delta^{2}(x^{A}-x^{A\prime}),
	\end{equation}
	where $\sigma=\pm$ indexes the whether the function lives on the future/past
	portion of the boundary, and where $q$ is the determinant of the metric on the
	unit two-sphere.  Using this we can see that the gauge invariant amplitude
	satisfies
	
	\begin{equation}
		-i\frac{\delta}{\delta
			a^{\sigma}_{B}(x^{A})}K(A_{\partial\mathcal{V}})=\bigg(\int_{\partial\mathcal{V}^{\sigma}}
		d^{3}x'\sqrt{g}J^{\eta}(x')\nabla^{B}\bar{G}(x^{A\prime},x^{A})\bigg)K(A_{\partial\mathcal{V}}),
	\end{equation}
	on the each of the future/part portions of the causal diamond boundary.
	
	It remains to understand what, physically, this functional differential
	operator represents. We can do so by using the relationship between functional
	derivatives and the symplectic current density in
	(\ref{eq:symplecticcurrentdensity}). For the gauge field $A_{B}$ we have
	\begin{equation}
		\theta^{\eta}(A_{\mu},\delta
		A_{B})=\frac{\partial\mathcal{L}}{\partial\nabla_{\eta}A_{B}}\delta
		A_{B}=-\sqrt{g}F^{\eta B}\delta A_{B},
	\end{equation}
	and if we separate the field as $A_{B}=\bar{A}_{B}+a_{B}$, where $a_{B}$ is
	independent of $r$ and $\bar{A}_{B}$ is vanishing at spatial infinity, then by
	linearity we have
	\begin{equation}
		\theta^{\eta}(A_{\mu},\delta
		a_{B})=\frac{\partial\mathcal{L}}{\partial\nabla_{\eta}A_{B}}\delta
		a_{B}=-\sqrt{g}F^{\eta B}\delta a_{B}.
	\end{equation}
	
	Invoking (\ref{eq:symplecticcurrentdensity}) we then find
	\begin{align}
		-i\frac{\delta}{\delta a^{\sigma}_{B}(x^{A})}
		K(A_{\partial\mathcal{V}})&=\int_{A_{\partial\mathcal{V}}}\mathcal{D}A_{\mu}\,e^{iS[A]}\bigg(\int_{\partial\mathcal{V}^{\sigma}}d^{3}x'\,\sqrt{g}F^{B\eta}q^{-1/2}\delta^{2}(x^{A}-x^{A\prime})\bigg)
		\nonumber \\
		&=\int_{A_{\partial\mathcal{V}}}\mathcal{D}A_{\mu}\,e^{iS[A]}\bigg(\int_{0}^{\infty}dr\,r^{2}F^{B\eta}(r,x^{A})\bigg|_{\partial\mathcal{V}^{\sigma}}\bigg).
	\end{align}
	
	The bFP trick has then illustrated that physical (gauge invariant) states on
	the boundary of the large causal diamond satisfy the eigenvalue equation
	\begin{equation}\label{eq:softchargeannil}		
		\bigg(\int_{0}^{\infty}dr\,r^{2}\hat{F}^{B\eta}(r,x^{A})\bigg|_{\partial\mathcal{V}^{\sigma}}\bigg)K(A_{\partial\mathcal{V}})=\bigg(\int_{\partial\mathcal{V}^{\sigma}}
		d^{3}x'\sqrt{g}J^{\eta}(x')\nabla^{B}\bar{G}(x^{A\prime},x^{A})\bigg)K(A_{\partial\mathcal{V}}),
	\end{equation}
	at every angle $x^{A}$ on the sphere, independently on each of the future
	and past parts of the boundary.  This is an exact relation, irrespective of the
	data specified for the fields or the dynamics of the charged matter, ie. it is
	kinematically required. It is a direct consequence of gauge invariance for the
	causal diamond path-integral on the extended configuration space when the gauge
	fields are allowed to take finite values at spatial infinity.
	
	This result bears a clear resemblance to results at null infinity which have
	been widely discussed in the literature \cite{CQG18,strominger,IRstuff}. Indeed,
	since (\ref{eq:softchargeannil}) holds at each angle, we can multiply it by
	$\partial_{B}\varepsilon(x^{A})$, for any function on the sphere
	$\varepsilon(x^{A})$, and integrate over the sphere. We then obtain
	\begin{equation}
		\int_{\partial\mathcal{V}^{\sigma}}
		d^{3}x\,\sqrt{q}\bigg((q^{AB}\nabla_{A}\varepsilon(x^{A}))\hat{F}_{Br}+\varepsilon(x^{A})
		(r^{2}J_{r})\bigg)K(A_{\partial\mathcal{V}})=0.
	\end{equation}
	
	If we go to complex stereographic coordinates $(z,\bar{z})$ on the unit
	sphere such that the metric is
	\begin{equation}
		d\Omega^{2}=2\gamma_{z\bar{z}}dzd\bar{z},
	\end{equation}
	with
	\begin{equation}
		\gamma_{z\bar{z}}=\frac{2}{(1+z\bar{z})^{2}},
	\end{equation}
	we obtain the operator constraint equation
	\begin{equation}\label{eq:largegaugecharge}
		\int_{\partial\mathcal{V}^{\sigma}}
		drd^{2}z\,\bigg(-\partial_{z}\varepsilon(z,\bar{z})\hat{F}_{\bar{z}r}-\partial_{\bar{z}}\varepsilon(z,\bar{z})\hat{F}_{zr}+\varepsilon(z,\bar{z})
		\gamma_{z\bar{z}}(r^{2}J_{r})\bigg)K(A_{\partial\mathcal{V}})=0.
	\end{equation}

	If we recall that, in our coordinates, $r$ is the affine parameter
	on the null boundary, then we immediately recognize the operator above, acting 
	on $K(A_{\partial\mathcal{V}})$, as the large gauge charge operator 
	$\hat{Q}_{\varepsilon}$ discussed in the recent literature (compare ref. 
	\cite{strominger}, section 2.5.11, and refs. therein). Thus, writing this 
	constraint equation as 
	\begin{equation}\label{eq:asymchargeannil}
		\hat{Q}_{\epsilon}^{\pm}K(A_{\partial\mathcal{V}})=0,
	\end{equation}
	we can say that (i) the electric part of 
	this operator is the zero-frequency part of the leading $\mathcal{O}(r^{0})$ 
	transverse electric field at infinity, and it thus creates soft photon states; 
	and (ii) the matter term is the total flux through the null surface of the 
	leading $\mathcal{O}(r^{-2})$ term in the matter current at infinity.

	When the matter is quantum mechanical the computation can be carried though 
	with no additional complications, and the result is to simply replace $J_{r}$ 
	by a functional differential operator representation of the $U(1)$ current 
	operator $\hat{J}_{r}$. Thus, the amplitude we've derived is annihilated by 
	the large gauge asymptotic charges $\hat{Q}_{\epsilon}^{\pm}$, for all 
	functions $\epsilon(z,\bar{z})$ on both the future and past surface.
	
	From our analysis we can see that since we've blown up spatial infinity,
	thereby allowing the value of the gauge field at spatial infinity to be
	different depending on whether approached from the future or past part of the 
	boundary, the amplitude satisfies (\ref{eq:largegaugecharge}) on the past and 
	future null boundaries separately. As a consequence, the states on the past 
	and future parts are necessarily dressed by soft photons in the way described 
	originally by Kibble, Chung, and Faddeev and Kulish 
	\cite{kibbleIR,chung,faddeevK} in such a way as to render scattering 
	amplitudes infrared-finite. This is clear from the form of
	eq. (\ref{eq:largegaugecharge}), which demonstrates that a non-zero flux of
	charge through either of the null surfaces is necessarily accompanied by
	infinite wavelength electric field excitations. The form of the dressing can
	also be seen explicitly in eq. (\ref{eq:largediamondampl}), where the 
	exponentiated positive (negative) frequency part of the gauge field on
	the past (future) null surface explicitly describes a coherent state sourced by
	the total charge flux through each angle.
	
	Let us relate this result further to the commonly used notation. To be 
	concrete,  we consider just the future part of the boundary, and confine 
	ourselves to massless charged particles. In comparing our coordinates with 
	the typical null coordinates it is clear that our $r$ is simply (up to an 
	irrelevant constant shift), equal to $-u$, where $u$ is the retarded time. In terms of the retarded
	time and stereographic angular coordinates, eq. (\ref{eq:softchargeannil}) is
	written (now with all lowered indices) as
	\begin{equation}\label{eq:softchargeannil2}		
		\bigg(\int_{-\infty}^{\infty}du\,\hat{F}^{(0)}_{zu}(u,z,\bar{z})-\int_{-\infty}^{\infty}
		du\,\int
		d^{2}\omega\,\gamma_{\omega\bar{\omega}}J^{(2)}_{u}(u,\omega,\bar{\omega})\partial_{z}\bar{G}(z,\bar{z},\omega,\bar{\omega})\bigg)K(A_{\partial\mathcal{V}})=0,
	\end{equation}
	where we now use the superscripts $(n)$ to denote the leading
	$\mathcal{O}(r^{-n})$ part of the operator. The charge density describing a
	collection of particles, each with charge $Q_{k}$, reaching the future null
	surface at $(u_{k},z_{k},\bar{z}_{k})$, is given by
	\begin{equation}\label{eq:chargeflux2}		
		J^{(2)}_{u}(u,z,\bar{z})=\sum_{k=1}^{m}Q_{k}\delta(u-u_{k})\gamma^{z\bar{z}}\delta^{(2)}(z-z_{k}).
	\end{equation}
	The last element needed to complete the translation is the expression for
	Green's function on the two-sphere, viz.,
	\begin{equation}\label{eq:stereogreens}
		\bar{G}(z,\bar{z},\omega,\omega)=\frac{1}{2\pi}\ln |z-\omega|^{2}.
	\end{equation}
	Inserting eqs. (\ref{eq:chargeflux2},\ref{eq:stereogreens}) into eq.
	(\ref{eq:softchargeannil2}) we then obtain
	\begin{equation}\label{eq:softchargeannil3}		
		\bigg(\int_{-\infty}^{\infty}du\,\hat{F}^{(0)}_{uz}(u,z,\bar{z})+\sum_{k\,\in \,
			\textrm{out}}\frac{Q_{k}}{2\pi}\frac{1}{z-z_{k}}\bigg)K(A_{\partial\mathcal{V}})=0.
	\end{equation}
	A completely analogous equation also holds on the past boundary.

	The first term in eq. (\ref{eq:softchargeannil3}) is often referred to as
	the soft-photon mode, $N^{+}_{z}$, while the second term is referred to as the
	soft-factor, $\Omega^{soft +}$ (for example, see
	\cite{strominger,newsymmetries2014,kapec2017} and refs. therein). We can then 
	express our result concisely as
	\begin{equation}\label{eq:chargevanish}
		(N_{z}^{+}+\Omega^{soft+})K(A_{\partial\mathcal{V}})=	
		(N_{z}^{-}+\Omega^{soft-})K(A_{\partial\mathcal{V}})=0,
	\end{equation}
	which is simply a restatement of eq. {\ref{eq:asymchargeannil}} for the
	particular choice $\varepsilon=(z-\omega)^{-1}$.  We have thus demonstrated that
	the physical states described by the propagator (\ref{eq:largediamondampl}) have
	strictly zero charge under large gauge transformations. In a sense this result
	is unsurprising, given that we have started from an expression (\ref{eq:amp4.2})
	which was manifestly invariant under all gauge transformations. However the main
	result, eq. (\ref{eq:largediamondampl}), is interesting in that it explicitly
	demonstrates how the soft-photon dressing emerges naturally as a consequence of
	gauge invariance.
	
	The relationship between the Faddeev-Kulish dressing factor and large gauge
	transformations has been previously been expressed clearly in the context of
	gravitational scattering by Choi and Akhoury \cite{akhoury2018}. They have
	demonstrated that scattering amplitudes between states with non-zero, but
	conserved, asymptotic charge are the Faddeev-Kulish dressed states.
	Additionally, they illustrated that there is considerable freedom in defining
	such states, allowing one to shuffle soft-photon contributions between the
	future and past states. Our result in eqs. (\ref{eq:largediamondampl},
	\ref{eq:chargevanish}) is consistent with this statement; however it is more 
	restrictive. The asymptotic charge is indeed conserved during the
	evolution; however we've also found that it is necessarily zero, and there is no
	freedom in shuffling around the soft-photon parts---both initial and final
	states of charged particles must have accompanying soft-photon dressing.
	
	Let us understand why we have arrived at this more stringent condition. The
	language parallels the discussion at the end of Sec. (\ref{sec:largeG-TS}). In
	the above calculations the asymptotic part of the gauge field, $a_{B}$, was
	allowed to be different on the past and future surfaces, and as a result so too
	were the large gauge transformations. The amplitude was thus invariant under the
	group $U(1)^{+}_{z\bar{z}}\times U(1)^{-}_{z\bar{z}}$, of angle dependent $U(1)$
	transformations which act independently on the past and future surfaces, and
	thus the initial and final states were independently dressed.
	
	This should be contrasted with the conjectured antipodal ``matching
	condition'' between fields at $\mathcal{I}^{+}_{-}$ and $\mathcal{I}^{-}_{+}$, 
	which would imply that the asymptotic gauge symmetry is only the diagonal
	subgroup $U(1)^{0}_{z\bar{z}}$ which acts simultaneously on both future and
	past~\cite{strominger2014}~\cite{newsymmetries2014,strominger2014}. The Ward
	identity associated with this symmetry was demonstrated to be equivalent to the
	soft photon theorem~\cite{low1958, weinberg89,newsymmetries2014}, providing
	further evidence that the matching condition is correct.  Furthermore, the
	conformal embedding of Minkowski spacetime in the Einstein static cylinder is
	smooth at spatial infinity $i^{0}$, and provides a natural antipodal
	relationship between the null generators of $\mathcal{I}^{+}$ and
	$\mathcal{I}^{-}$. In flat spacetime it is then natural to assume the matching
	condition. However for more general spacetimes the requisite smoothness at
	$i^{0}$ does not hold, and the matching problem has long been an open
	question~\cite{ashtekarhansen,ashtekartalk}. For this reason, to allow for a
	generalization to gravitational theories in future work, we have explicitly
	blown up the region analogous to $i^{0}$ in this paper, and enforced no matching
	condition\footnote{During peer review we became aware of several recent
		works~\cite{prabhupapers} which prove that the conjectured matching condition
		will hold, for asymptotic symmetry charges, in a rigorously defined general
		class of asymptotically flat spacetimes. This then suggests that one should
		gauge fix $\nabla^{B}a_{B}$ on only one of the future and past surfaces.}

	\subsubsection{Enforcing the Antipodal Matching Condition}
	Let us now complete the discussion by assuming that the antipodal matching
	condition holds, and repeat the calculation. We then impose
	\begin{equation}
		a_{B}^{+}(z,\bar{z})=a_{B}^{-}(z,\bar{z}),
	\end{equation}
	where we've used the antipodal mapping $z\rightarrow-\tfrac{1}{\bar{z}}$ between
	the coordinates on the future surface relative to those on the
	past~\cite{strominger}. There are two main technical differences in the
	calculation, viz.:
	\begin{itemize}
		\item when modifying the gauge fixing procedure (\ref{eq:largeGFP}) we now
		need only to fix
		\begin{equation}
			\nabla^{B}a_{B}^{+}(z,\bar{z})=0,
		\end{equation}
		\item the functional derivative of the symplectic current density in eq.
		(\ref{eq:symplecticcurrentdensity}) now has support on both parts of the
		boundary, causing the conjugate momentum to $a_{B}$ to be the difference
		between photon zero modes,
		\begin{equation}
			-i\frac{\delta}{\delta a_{B}(x^{A})}		
			\dot{=}\int_{0}^{\infty}dr\,r^{2}F^{B\eta}(r,x^{A})\bigg|_{\partial\mathcal{V}^{+}}-\int_{0}^{\infty}dr\,r^{2}F^{B\eta}(r,x^{A})\bigg|_{\partial\mathcal{V}^{-}}.
		\end{equation}
	\end{itemize}
	The remaining steps in the calculation are identical, and we arrive at the final
	expression analogous to eq. (\ref{eq:largediamondampl})
	\begin{align}\label{eq:largediamondampl2}
		K(A_{\partial\mathcal{V}})& \;\;=\;\;
		\delta\left(\int_{\partial\mathcal{V}}J^{\eta}\right)\exp\left[-\frac{i}{2\pi}\int
		d^{2}z\, a_{z}(z,\bar{z})\int_{\partial\mathcal{V}}d\lambda\,
		\frac{J_{\lambda}^{(2)}(\lambda,\omega,\bar{\omega})}{z-\omega}-h.c.\right]
		\nonumber \\	
		&e^{i\int_{\partial\mathcal{V}}d^{3}x\sqrt{g}J^{\eta}\big[\int_{\partial\mathcal{V}}d^{3}x'\sqrt{g}GD^{j}A_{j}\big]}\times
		\int_{\mathcal{A}_{j\,\partial\mathcal{V}}}\mathcal{D}\bar{A}_{\eta}\mathcal{D}A_{j}\,\delta(D^{j}A_{j})\,e^{iS[A_{j},\bar{A}_{\eta}]}.
	\end{align}
	where $\lambda$ is the affine parameter along each null surface. We can now
	explicitly see that the dressing involves the difference between charge fluxes
	through each surface. Furthermore, by functionally differentiating with respect
	to $a_{z}$ we can obtain the analogous expression to eqs.
	(\ref{eq:softchargeannil2}, \ref{eq:softchargeannil3}), ie.,
	\begin{equation}	
		(N_{z}^{+}-N_{z}^{-})K(A_{\partial\mathcal{V}})=(\Omega^{soft-}-\Omega^{soft+})K(A_{\partial\mathcal{V}}).
	\end{equation}
	These two terms need not vanish, and so we have recovered the asymptotic charge
	conservation law of ref.~\cite{newsymmetries2014}, as we expected.

	
	\section{Conclusions}
	\label{sec:conclusion}
	
	
	Let us begin here by summarizing what has been done, and then return to the
	questions asked in the introduction.
	
	This paper has defined a manifestly gauge invariant analysis of QED
	amplitudes and quantum states for flat space QED. We use the modern
	understanding of states as data living on the closed hypersurface boundaries of
	path-integrals, in a `general boundary QFT framework', in which the
	path-integral allows us to go beyond canonical quantization. The interpretation
	in terms of states and transition amplitudes is then secondary, and only applies
	to particular geometries.
	
	The bFP trick discussed here should be applicable for general boundaries. In
	the 2 cases considered here, we found
	
	(a) when $\partial\mathcal{V}$ consists of two finitely separated constant
	time slices and a time-like cylinder at spatial infinity, we can interpret
	$K_{fi}$ as a conventional transition amplitude which also has a representation
	in terms of a Hamiltonian operator;
	
	(b) when $\partial\mathcal{V}$ is the null boundary of a large causal
	diamond, $K_{fi}$ is most naturally described as the path-integral. The causal
	diamond boundary then resembles null infinity, and the amplitude resembles a
	scattering amplitude.
	
	As a consequence of the gauge invariance of the QED action, these amplitudes
	were gauge-invariant and independent of non-canonical variables, and were
	written explicitly in terms of gauge invariant variables. One obtains unique
	expressions for the dependence of the amplitudes on the gauge-variant parts of
	$A_{\mu}$; this dependence arose only as a boundary phase.
	
	Thus, rather than solving the constraint equation (which under-determines
	the state), we found that the path integral yields unique expressions for the
	boundary phases for which the constraint equation is also satisfied. The
	dependence of $K_{fi}$ on the gauge-variant parts of the field was determined
	kinematically, whereas the dependence on the gauge-invariant parts of the field
	was found dynamically, by a path integral over gauge invariant variables.
	
	In cases when gauge transformations vanished at spatial infinity one
	inevitably finds a universal Coulombic dressing of the charges in the boundary
	state, as well as some non-universal, dynamically generated,
	transverse contributions to the dressing.  However when the gauge group was
	extended, the causal diamond amplitude boundary states were annihilated by the
	``large-gauge charge'' discussed previously in the literature on null infinity
	\cite{largeGT,strominger}. One can explicitly enforce an asymptotic
	matching condition at spatial infinity and demonstrate that this zero-charge
	condition weakens to  the asymptotic charge conservation laws associated with
	the soft photon theorem.  From the large causal diamond path integral an
	explicit expression emerges very naturally for the soft-photon dressing of
	states on the null boundary. The resulting expressions were not novel, but the
	bFP technique used to get them was, and it provided a manifestly gauge invariant
	derivation of the result.
	
	Let us now turn to the questions posed in the introduction. From the
	foregoing discussion, we can see that a fairly complete characterization has
	been given here of the states, their dressing, the physical degrees of freedom,
	and how all these depend on the boundaries (questions (a)-(c) of the
	introduction).
	
	Let us then turn to question (d), regarding the larger enterprise of quantum
	gravity. Here we can only indicate which direction we believe should be taken -
	to really deal with this, one needs to generalize the work here to much more
	general spacetime structures. For simple spacetimes, one can certainly extend
	the class of allowed gauge transformations beyond those considered here - gauge
	invariance would then imply new angle dependent constraints on our asymptotic
	states, perhaps related to the sub-leading soft photon theorems.
	
	We note however that the methods developed here will allow discussion of a
	key question in current debates about quantum gravity, viz., whether the metric
	field really needs to be quantized at all, and how one might test this question
	experimentally, in, eg., ``BMV" experiments
	\cite{vedral17,milburn17,beilok18,belenchia18}. The answer to this turns
	essentially on how one defines physical states for the metric field. The
	generalization of our methods to linearized gravity - which is all that is
	necessary to deal with this problem - is straightforward, and will also be
	discussed elsewhere.
	
	It is also of interest to try and define states using path integrals in more
	general theories of quantum gravity. An example is provided by the CWL theory
	\cite{stampCWL,CWL2,CWL3}, in which QM breaks down because of gravitational
	correlations between paths. Very recently we have shown how both standard 2-path
	superposition experiments and ``4-path" experiments of the BMV type can be
	analyzed \cite{Jordan-PCES} in CWL theory - this theory in fact lends itself
	very naturally to the boundary quantum field theory framework which underpins
	the analysis of the present paper.

	
	\section{Acknowledgements}
	\label{sec:Ack}
	
	
	This work was funded in Canada by the National Science and Engineering
	Research Council of Canada (NSERC). In the USA, PCES received support at Caltech
	from the Simons Foundation (Award 568762) and the National Science Foundation
	(Award PHY-1733907). JWG was additionally supported by a NSERC PGS-D award in
	Canada, and a Burke Fellowship at Caltech.
	
	We have benefited particularly from extended discussions with C Delisle at
	UBC. We also thank W.G. Unruh at UBC and Y. Chen at Caltech for illuminating
	remarks. We would also like to acknowledge the support and hospitality of Y.
	Chen, T.F. Rosenbaum, and K.S. Thorne at Caltech.

	\appendix
	
	
	\section{Gauge Transformations on $K_{fi}$}
	\label{sec:App-A}
	
	
	Here we give some of the formal manipulations involved in deriving the key
	expressions in section 2, for the gauge-invariant propagator $K_{fi}$ defined
	between 2 time slices in flat spacetime. This is done first for scalar
	electrodynamics, as defined by the action given in eqns.     ,and then for
	spinor QED, defined by the action in eqn.

	\subsection{Scalar Electrodynamics}
	\label{sec:A-scalarQED}
	
	In the main text we began from eqn. (\ref{eq:particlepropagator}) for the
	combined matter/EM field propagator for scalar electrodynamics, which we repeat
	here;
	\begin{eqnarray}
		\label{eq:particlepropagator2a}
		K_{fi} & \equiv & K(q_{f},A_{\mu\,f};q_{i},A_{\mu\,i}) \nonumber \\
		&=&
		\int^{q_{f}}_{q_{i}}\mathcal{D}q\,e^{iS_{M}}\int^{A_{\mu\,f}}_{{A_{\mu\,i}}}\mathcal{D}A_{\mu}\,e^{iS_{EM}}.
	\end{eqnarray}
	which we argued could also be written in the form of eqn.
	(\ref{eq:finalprop}), which we also repeat here:
	\begin{equation}
		\label{eq:finalprop2a'}
		K_{fi} = e^{i\tilde{S}_C}  \int^{q_{f}}_{q_{i}}\mathcal{D}q\,e^{i
			\tilde{S}_{M}}  \int^{\mathcal{A}^j_f}_{\mathcal{A}^j_i} \mathcal{D}
		\mathcal{A}^{j}\,e^{i \tilde{S}_A}.
	\end{equation}
	
	In what follows we will first demonstrate explicitly the gauge invariance of
	(\ref{eq:particlepropagator2a}), and then give details of the derivation of
	(\ref{eq:finalprop2a'}), using the two different methods described in section
	2.A.

	\subsubsection{Gauge Invariant Propagator}
	\label{sec:gaugeI-K}
	
	In the main text we argued that eqn. (\ref{eq:particlepropagator}) for the
	combined propagator was manifestly gauge invariant. Here we amplify on this
	assertion, and demonstrate it explicitly. Under gauge transformation eqn. became
	(recall eqn. (\ref{K-fi})):
	\begin{eqnarray}
		K_{fi}^{\Lambda} &\equiv &
		K^{\Lambda}(q_{f},A_{\mu\,f};q_{i},A_{\mu\,i}) \nonumber \\		
		&=&e^{-ie\Lambda_{f}(q_{f})}K(q_{f},A^{\Lambda_{f}}_{\mu\,f};q_{i},A^{\Lambda_{i}}_{\mu\,i})e^{ie\Lambda_{i}(q_{i})}
		\label{K-fi2}
	\end{eqnarray}
	
	Now this propagator, with transformed boundary data, can be expressed simply
	in terms of the original propagator. We take the expression
	\begin{equation}		
		K(q_{f},A^{\Lambda_{f}}_{\mu\,f};q_{i},A^{\Lambda_{i}}_{\mu\,i})=\int^{q_{f}}_{q_{i}}\mathcal{D}q\,e^{iS_{M}}\int^{A^{\Lambda_{f}}_{\mu\,f}}_{{A^{\Lambda_{i}}_{\mu\,i}}}\mathcal{D}A_{\mu}\,e^{iS_{EM}}
	\end{equation}
	and perform a change of variables, $A_{\mu}=A'_{\mu}+\partial_{\mu}\Lambda$,
	for some time-dependent function $\Lambda$ which takes the value $\Lambda_{i,f}$
	on $\Sigma_{i,f}$. The boundary data for the new variable $A'_{\mu}$ is now just
	the original configuration, $A_{\mu\,i,f}$. The action is not invariant under
	this change of variables, but instead, since it is effectively just a gauge
	transformation, we know that the action changes by a simple boundary term. The
	action is expressed in terms of $A'_{\mu}$ as
	\begin{equation}\label{eq:deltaaction}
		S_{EM}[q,A]=S_{EM}[q,A']+e\Lambda_{f}(q_{f})-e\Lambda_{i}(q_{i}),
	\end{equation}
	so that the propagator with transformed boundary data then reads
	
	\begin{align}
		\label{eq:transbndry}		
		K(q_{f},A^{\Lambda_{f}}_{\mu\,f};q_{i},A^{\Lambda_{i}}_{\mu\,i})&=\int^{q_{f}}_{q_{i}}\mathcal{D}q\,e^{iS_{M}[q]}\int^{A_{\mu\,f}}_{{A_{\mu\,i}}}\mathcal{D}A'_{\mu}\,e^{iS_{EM}[q,A']+e\Lambda_{f}(q_{f})-e\Lambda_{i}(q_{i})}
		\nonumber \\
		&=e^{ie\Lambda_{f}(q_{f})} \; K_{fi} \; e^{-ie\Lambda_{i}(q_{i})}.
	\end{align}
	
	The boundary phases in (\ref{eq:transbndry}) generated by the gauge-field
	action will then precisely cancel the phases in (\ref{K-fi2}) arising from the
	$U(1)$ transformation of the matter states, and therefore the propagator for the
	total system is gauge invariant.

	\subsubsection{Constraint Equation}
	\label{sec:constraintEq}
	
	Here we describe the derivation of eqn. (\ref{constraint1}) from eqn.
	(\ref{K-fi}) in the main text.  To do this, consider a gauge transformation
	which vanishes on $\Sigma_{i}$ but not on $\Sigma_{f}$, and rewrite the
	transformed propagator (\ref{K-fi}) using a linear shift operator as
	\begin{align}
		\label{eq:linearshiftb}
		K^{\Lambda}_{fi}\;=\; e^{-ie\Lambda_{f}(q_{f})+\int_{\Sigma_{f}}
			d^{3}x\partial_{\mu}\Lambda_{f}\frac{\delta}{\delta A_{\mu\,f}}} \; K_{fi}
	\end{align}
	
	Since the propagator is gauge invariant, this implies the following simple
	functional differential equation
	
	\begin{align}
		0& \;\;=\;\;\bigg[-ie\Lambda_{f}(q_{f})+\int_{\Sigma_{f}}
		d^{3}x\partial_{\mu}\Lambda_{f}\frac{\delta}{\delta A_{\mu\,f}}\bigg] \; K_{fi}
		\nonumber \\
		&\;\;=\;\; \bigg[\int_{\Sigma_{f}}
		d^{3}x\,\partial_{0}\Lambda_{f}\frac{\delta}{\delta A_{0\,f}}-i\int_{\Sigma_{f}}
		d^{3}x\Lambda_{f}\bigg(e\delta^{3}(q_{f}-x)-i\partial_{j}\frac{\delta}{\delta
			A_{j\,f}}\bigg)\bigg] \; K_{fi}
	\end{align}

	The remaining functional derivative of the propagator with respect to
	$A_{j\,f}$ is just the electric field operator. This is seen in a path integral
	treatment by evaluating the functional derivative and using the standard
	expression for variations of the action endpoint in mechanics, viz.,
	\begin{equation}
		\frac{\delta S[x]}{\delta
			x_{f}}=\frac{\partial\mathscr{L}}{\partial\dot{x}(t)}\bigg|_{t_{f}}.
	\end{equation}
	
	Written in terms of the electric field operator we then get the constraint
	equation
	\begin{eqnarray}
		0 &=&\bigg[\int d^{3}x\,\partial_{0}\Lambda_{f}\frac{\delta}{\delta
			A_{0\,f}} \nonumber \\
		&& \;\;\; -i\int
		d^{3}x\Lambda_{f}\bigg(e\delta^{3}(q_{f}-x)-\partial_{j}\hat{E}^{j}\bigg)\bigg]
		\; K_{fi} \qquad
	\end{eqnarray}
	which is just eqn. (\ref{constraint1}) of the main text.

	\subsection{Extracting the Dressing}
	
	Here we give details of (i) the integration over $A_0$ appearing in eqn.
	(\ref{eq:particlepropagator2}) of the main text, and (ii) the derivation of the
	final form (\ref{eq:finalprop}) for the scalar QED propagator $K_{fi}$, along
	with the transformed form for the effective action which appears in it (ie.,
	eqn. (\ref{tilde-S}), containing the three terms
	(\ref{tildeSM})-(\ref{tildeSA})).

	\subsubsection{Integration over $A_0$}
	\label{sec:A0-int}
	
	We recall eqn. (\ref{eq:particlepropagator2}) of the main text for the
	propagator $K_{fi}$, which we repeat here:
	\begin{equation}
		\label{eq:particlepropagator2'}
		K_{fi} \;=\;
		\int^{q_{f}}_{q_{i}}\mathcal{D}q\,e^{iS_{M}}\int\mathcal{D}A_{0}\int^{A_{j\,f}}_{{A_{j\,i}}}\mathcal{D}A_{j}\,e^{iS_{EM}}.
	\end{equation}
	We now start from (\ref{eq:particlepropagator2'}), and evaluate the $A_{0}$
	integral. No gauge-fixing is required to make this integral convergent, and the
	boundary data is unfixed, so we can directly go ahead and do the integration. As
	we will see, this rather uniquely determines how we should gauge fix the
	remaining $A_{j}$ integral, and consequentially determines the form of the
	dressing for the states.
	
	We first separate out the $A_{0}$ dependent terms in the electromagnetic
	part of the action:
	\begin{eqnarray}
		S_{EM}
		&=&\int^{t_{f}}_{t_{i}}d^{4}x\bigg[-\frac{1}{4}F_{jk}F^{jk}+A_{j}J^{j} \nonumber
		\\ && \qquad\qquad
		-\frac{1}{2}F^{j0}(\partial_{j}A_{0}-\partial_{0}A_{j})+A_{0}J^{0}\bigg]  \qquad
	\end{eqnarray}
	Integrating the spatial derivatives by parts we obtain
	\begin{eqnarray}
		\label{eq:act1}
		S_{EM} &=&
		\int^{t_{f}}_{t_{i}}d^{4}x\bigg[-\frac{1}{4}F_{jk}F^{jk}+A_{j}J^{j} \nonumber \\
		&+&\frac{1}{2}F^{j0}\partial_{0}A_{j}+\frac{1}{2}A_{0}\big(\partial_{j}F^{j0}+J^{0}\big)+\frac{1}{2}A_{0}J^{0}\bigg]
		\qquad
	\end{eqnarray}
	
	The variable $A_{0}$ appears quadratically in the action, and since its
	endpoints are being integrated over in (\ref{eq:particlepropagator2'}), it can
	be integrated out as a simple Gaussian integral. The result of course is to just
	substitute the saddle point solution $\tilde{A}_{0}$ back into (\ref{eq:act1}).
	
	The saddle point equation for $A_{0}$ is just the Gauss law Maxwell
	equation, viz.,
	\begin{equation}\label{eq:gausslawmaxwelleqn}
		(\partial_{j}F^{j0}+J^{0}) \;=\;
		-\partial_{j}\partial^{j}A_{0}+\partial_{0}\partial^{j}A_{j}+J^{0} \;=\; 0.
	\end{equation}
	for which the solution is
	\begin{equation}\label{eq:A0tilde}
		\tilde{\mathcal{A}}_{0}=\nabla^{-2}J^{0}+g+h,
	\end{equation}
	where $g$ is given by
	\begin{equation}\label{eq:gflat}
		g=\partial_{0}\nabla^{-2}(\partial^{j}A_{j}),
	\end{equation}
	and where $h$ is an undetermined homogeneous solution to the Laplace
	equation. The only such solution which is both regular at the origin and
	vanishes at spatial infinity is the trivial solution, $h(x)=0$, so we set $h=0$.
	
	The solution (\ref{eq:A0tilde}) is then unique, without needing to impose
	further gauge fixing to eliminate the homogeneous solutions (if however we allow
	for large gauge transformations, then non-trivial expressions for $h(x)$ arise,
	and further gauge fixing is required - see section 4 of the main text).
	
	Notice that $\tilde{A}_{0}$ is given in terms of a gauge invariant term
	$\nabla^{-2}J^{0}$ and a gauge variant term $g$. Under gauge transformation $g$
	transforms as $\delta_{\Lambda}g=\partial_{0}\Lambda$, as it must, so that
	$\tilde{A}_{0}$ transforms appropriately.
	
	Taking inspiration from this, we formally isolate the gauge invariant part
	of the components $A_{j}$ by defining
	\begin{equation}\label{eq:ajdecomp}
		A_{j}=\mathcal{A}_{j}+\partial_{j}\Phi,
	\end{equation}
	where $\delta_{\Lambda}\mathcal{A}_{j}=0$, and $\Phi$ is a functional of
	$A_{j}$ with the assumed transformation property $\delta_{\Lambda}\Phi=\Lambda$.
	The functions $(\mathcal{A}_{j},\Phi)$  are just a new choice of field variables
	for the path-integration. To avoid introducing a field-dependent Jacobian into
	the integration measure, we will assume the \textit{g-potential} $\Phi$ to be a
	linear functional of the $A_{j}$. Note that $\Phi$ is certainly not given
	uniquely by the required transformation property: for now we leave it
	unspecified.
	
	At this point one might assume that $\mathcal{A}_{j}$ and $\partial_{j}\Phi$
	are just the transverse and longitudinal parts of $A_{j}$. This is of course a
	valid decomposition, but not a unique one. We will instead rewrite the
	path-integral in terms of the new variables $\mathcal{A}_{j}$ and
	$\partial_{j}\Phi$, and look for a natural decomposition of the path integral.
	We will see that the action, transformed to the new variables, ends up
	separating into a non-dynamical boundary term, plus terms which are uniquely
	associated with the dynamical matter field and the new field variables
	$\mathcal{A}_{j}(x)$.

	\subsubsection{New Variables for the Action and Propagator}
	
	We begin by writing the propagator $K_{fi}$ in terms of $\mathcal{A}_j$ in
	(\ref{eq:ajdecomp}) and the solution (\ref{eq:A0tilde}) for $\tilde{A}_0$, to
	get
	\begin{equation}
		K_{fi} \;=\;
		\int^{q_{f}}_{q_{i}}\mathcal{D}q\,e^{iS_{M}}\int_{\Phi_{i}}^{\Phi_{f}}\mathcal{D}\Phi\int^{\mathcal{A}_{j\,f}}_{{\mathcal{A}_{j\,i}}}\mathcal{D}\mathcal{A}_{j}\,e^{i\tilde{S}_{EM}},
	\end{equation}
	with a new electromagnetic field action $\tilde{S}_{EM}$ given by
	
	\begin{equation}
		\tilde{S}_{EM} \;=\;
		\int_{t_{i}}^{t_{f}}d^{4}x\bigg[-\frac{1}{4}F_{jk}F^{jk}+\mathcal{A}_{j}J^{j}+\partial_{j}\Phi
		J^{j}+\frac{1}{2}\tilde{F}^{j0}\partial_{0}
		\mathcal{A}_{j}+\frac{1}{2}\tilde{F}^{j0}\partial_{0}\partial_{j}\Phi+\frac{1}{2}J^{0}\nabla^{-2}J^{0}+\tfrac{1}{2}gJ^{0}\bigg],
	\end{equation}
	where we've introduced the notation
	$\tilde{F}_{j0}=\partial_{j}\tilde{A}_{0}-\partial_{0}A_{j}$. Note that $F_{jk}$
	is independent of $\Phi$ by antisymmetry.
	
	We can now integrate by parts to strip the spatial derivatives off $\Phi$,
	to get
	\begin{equation}\label{eq:effaction}
		\tilde{S}_{EM} \;=\;
		\int_{t_{i}}^{t_{f}}d^{4}x\bigg[\frac{1}{2}\tilde{F}^{j0}\partial_{0}\mathcal{A}_{j}-\frac{1}{4}F_{jk}F^{jk}+
		\mathcal{A}_{j}J^{j}+\frac{1}{2}J^{0}\nabla^{-2}J^{0}-\Phi
		\partial_{j}J^{j}-\frac{1}{2}\partial_{j}\tilde{F}^{j0}\partial_{0}\Phi+\tfrac{1}{2}gJ^{0}\bigg].
	\end{equation}
	
	We then use the definition $\partial_{j}\tilde{F}^{j0}=-J^{0}$ along with
	the fact that $\partial_{\mu}J^{\mu}$ for an arbitrary trajectory of the
	particle, to rewrite the action as
	\begin{equation}\label{eq:presumderiv}
		\tilde{S}_{EM} \;=\;
		\int_{t_{i}}^{t_{f}}d^{4}x\bigg[\frac{1}{2}\tilde{F}^{j0}\partial_{0}\mathcal{A}_{j}-\frac{1}{4}F_{jk}F^{jk}+\mathcal{A}_{j}J^{j}+\frac{1}{2}J^{0}\nabla^{-2}J^{0}+\Phi
		\partial_{0}J^{0}+\frac{1}{2}J^{0}\partial_{0}\Phi+\tfrac{1}{2}gJ^{0}\bigg],
	\end{equation}

	This result reveals something remarkable---if we now make the choice
	$\partial_{0}\Phi=g$ for $\Phi$,  then the last three terms sum to a total time
	derivative. There is of course nothing forcing us to choose this form for
	$\Phi$; since we are just making a change of path-integration variable, the
	final result for the propagator cannot depend on which decomposition we choose
	and it is best to make the simplest choice, especially in the later
	sections with lengthier calculations \footnote{One can directly check
		that the result does not depend on this choice by noting that
		$g=\partial_{0}\nabla^{-2}(\partial^{j}A_{j})=\partial_{0}\nabla^{-2}(\partial^{j}\mathcal{A}_{j})+\partial_{0}\Phi$
		and observing that terms involving $\Phi$ always sum to total derivatives. We
		thank our referee for making this clear.}. We will make this choice, and since
	$g$ itself is given as the time derivative of $\nabla^{-2}(\partial_{j}A^{j})$,
	we can simply choose
	\begin{equation}
		\Phi=\nabla^{-2}(\partial_{j}A^{j}).
	\end{equation}
	so that our new field variable now becomes
	\begin{equation}
		\mathcal{A}_{j}=A_{j}-\partial_{j}\nabla^{-2}(\partial^{k}A_{k})
		\label{tildeA2}
	\end{equation}
	
	We will find this result ultimately corresponds to the
	transverse-longitudinal decomposition of $A_{j}$; however, instead of assuming
	this from the start, we will see that this decomposition is simply dictated by
	the solution to the $A_{0}$ saddle point equation. This pattern of logic will be
	used again in later sections when we consider geometries for which it is much
	less clear {\it a priori} how to define a `transverse part' of $A_{j}$.
	
	Let us now complete the process of transforming to the new form for the
	field action.
	Note first that our choice of decomposition also simplifies the expression
	for the electric field, to
	\begin{align}
		\tilde{F}_{j0}&=\partial_{j}\tilde{A}_{0}-\partial_{0}A_{j} \nonumber \\
		&=\partial_{j}\nabla^{-2}J^{0}-\partial_{0}\mathcal{A}_{j},
	\end{align}
	ie., a manifestly gauge invariant form; and it renders $\mathcal{A}_{j}$
	divergenceless.
	
	We then find that a simple integration by parts gives
	\begin{equation}		
		\int_{t_{i}}^{t_{f}}d^{4}x\frac{1}{2}\tilde{F}^{j0}\partial_{0}\mathcal{A}_{j}=\int_{t_{i}}^{t_{f}}d^{4}x\frac{1}{2}\partial_{0}\mathcal{A}^{j}\partial_{0}\mathcal{A}_{j}.
	\end{equation}
	so that the field action takes the form
	\begin{eqnarray}
		&&\tilde{S}_{EM} \;=\; \int_{\partial\mathcal{V}} \sigma \;
		d^{3}xJ^{0}\nabla^{-2}(\partial_{j}A^{j}) + \tfrac{1}{2}
		\int_{t_{i}}^{t_{f}}d^{4}x\bigg[-\partial_{\mu}\mathcal{A}^{j}\partial^{\mu}
		\mathcal{A}_{j}+ 2\mathcal{A}_{j}J^{j}+J^{0}\nabla^{-2}J^{0}\bigg] \qquad
	\end{eqnarray}
	with $\sigma=\pm1$ for the future and past parts of the boundary
	respectively.
	
	We can now combine this field action with the original matter action $S_M$
	in (\ref{S-M}), to get a complete form for the transformed action, as
	\begin{equation}
		\tilde{S} \;=\; \tilde{S}_M + \tilde{S}_C + \tilde{S}_A
		\label{tilde-S'}
	\end{equation}
	which is eqn. (\ref{tilde-S}) of the main text, with the three new terms
	defined in  eqns. (\ref{tildeSM})-(\ref{tildeSA}).
	
	This leads finally to the propagator $K_{fi}$ in the form that we want.
	Since $\Phi$ does not appear in the bulk action, we can freely integrate over it
	to yield a harmless overall (divergent) normalization. Doing this, and
	continuing to absorb field independent constants into the measure, we arrive at
	eqn. (\ref{eq:finalprop}) of the main text, for $K_{fi}$. We see that all of the
	variables in the bulk action are gauge invariant, while the boundary term
	transforms precisely as we determined it ought to in (\ref{eq:transbndry}). Note
	also that the g-potential $\Phi$ is not present in the bulk action; it appears
	only in the boundary term.

	\subsection{Boundary Faddeev-Popov Trick}
	\label{sec:boundaryFP}
	
	Here we show that eqn. (\ref{eq:bFPstep3}) in the main text can be
	transformed into eqn. (\ref{Kfi-scalar-bFP}), after integration over $A_0$.
	
	We begin by noting that the $A_{0}$ integral can be performed unambiguously
	without need for gauge-fixing. We therefore assume a gauge-fixing function which
	does not involve $A_{0}$, and rewrite the transformed boundary data using a
	linear shift,  using functional derivatives as we did in
	(\ref{eq:linearshiftb}). We define the operator
	\begin{equation}
		\hat{\cal L}_{\Lambda} \;=\;
		\int_{\partial\mathcal{V}}d^{3}x\big[\Lambda
		J^{0}+i\partial_{\mu}\Lambda\frac{\delta}{\delta A_{\mu}} \big]
		\label{hatL}
	\end{equation}
	which now integrates over both past and future boundaries, and get
	\begin{align}
		\label{eq:bFPstep4}
		&K_{fi} \;=\;
		\int\mathcal{D}\Lambda\,\delta^{\partial\mathcal{V}}\big(\mathcal{G}(A^{\Lambda})\big)\,e^{-i
			\hat{\cal L}_{\Lambda}}   \int^{q_{f}}_{q_{i}}\mathcal{D}q\,e^{iS_{M}}
		\int^{A_{\mu\,f}}_{{A_{\mu\,i}}}\mathcal{D}A_{\mu}\,\Delta[A]\delta^{\mathcal{V}}\big(\mathcal{G}(A)\big)\,e^{iS_{EM}[A]}
		\qquad
	\end{align}
	
	The boundary delta function depends only on $\Lambda_{i,f}$ and not time
	derivatives thereof. The gauge transformations of the boundary data for $A_{0}$
	are then completely decoupled from the transformations of the remaining $A_{j}$.
	In a time-sliced discretization of the path integral, the transformation
	involving $\Lambda$ on the slices immediately after $\Sigma_{i}$ and
	$\Sigma_{f}$ will only affect the transformation of $A_{0}$ on the boundary.
	Additionally, there is no dependence in the integrand on $\Lambda$ for any
	intermediate times. This ``bulk'' integration over the gauge group can be
	factored out as usual, leaving a residual integration over boundary gauge
	transformations.
	
	The net result is that in (\ref{eq:bFPstep4}) we can rewrite $\hat{\cal
		L}_{\Lambda}$ as
	\begin{equation}
		\hat{\cal L}_{\Lambda} \;\; \rightarrow \;\;
		\int_{\partial\mathcal{V}}d^{3}x\left[\Lambda
		J^{0}+i\partial_{j}\Lambda\frac{\delta}{\delta A_{j}}\right]
		\label{hatL'}
	\end{equation}
	and omit the boundary data for $A_{0}$. The omission of $A_{0}$ boundary
	data dictates that its values on the boundary are integrated over.
	
	We can use the delta function to evaluate the integral over the boundary
	gauge transformations, and this will fix the boundary phase. Assuming
	$\mathcal{G}$ is a good gauge fixing function, it will correspond to a unique
	gauge parameter $\Lambda=\Lambda_{\mathcal{G}}[A]$. Evaluating the integral over
	the boundary gauge transformation we then obtain
	\begin{align}
		\label{eq:bFPchargedparticle}
		&K_{fi} \;=\; e^{-i \hat{\cal L}_{\Lambda_{\cal G}}}
		\int^{q_{f}}_{q_{i}}\mathcal{D}q\,e^{iS_{M}} \,
		\int^{A_{j\,f}}_{{A_{j\,i}}}\mathcal{D}A_{\mu}\,\Delta[A]\delta^{\mathcal{V}}\big(\mathcal{G}(A)\big)\,e^{iS_{EM}[A]}
		\qquad
	\end{align}
	where now
	\begin{equation}
		\hat{\cal L}_{\Lambda_{\cal G}} = \int_{\partial\mathcal{V}}d^{3}x \,
		\Lambda_{\cal G}[A] \left[ J^{0}+i\partial_{j}\Lambda\frac{\delta}{\delta
			A_{j}}\right]
		\label{hatL''}
	\end{equation}
	
	The difference between the bFP trick and the usual FP technique is clear
	from (\ref{eq:bFPchargedparticle}). While the path integral integrand itself is
	standard, the additional boundary phase effects a particular gauge
	transformation of the boundary data, which depends on the choice of bulk gauge
	fixing function $\mathcal{G}$. This boundary phase ensures that the resulting
	propagator remains independent of the choice of gauge fixing; it remains a gauge
	invariant object.
	
	Since the propagator is independent of gauge choice, we can choose the most
	convenient gauge. The argumentation is then similar to what we did earlier. We
	first recall that after the $A_{0}$ integration, and the change of variables to
	the invariant fields $\mathcal{A}_{j}$ and $\Phi$, we're left with an effective
	action (\ref{eq:effaction}). Great simplification came if we then chose
	$\partial_{0}\Phi=g$, where $g$ given in (\ref{eq:gflat}) was the unique
	gauge-dependent part of the saddle point solution $\tilde{A}_{0}$. Additionally,
	a few more terms in the effective action which involved $g$ and the current
	summed to a total derivative after using off-shell current conservation.
	
	We could actually skip the off-shell current conservation argument at this
	point, by simply choosing the Coulomb gauge $\mathcal{G}(A)=\partial^{j}A_{j}$.
	The particular usefulness of this gauge choice is that it sets $g=\Phi=0$,
	considerably simplifying the action. It also makes the FP determinant
	irrelevant, and implies
	\begin{equation}
		\Lambda_{\mathcal{G}}[A]=-\nabla^{-2}\partial^{j}A_{j},
		\label{LambdaG}
	\end{equation}
	for our boundary phases.
	
	The resulting expression for the propagator is
	\begin{align}
		&K_{fi} \;\;=\; \;
		e^{i\int_{\partial\mathcal{V}}d^{3}x\,\nabla^{-2}(\partial^{k}A_{k})\big[
			J^{0}-i\partial_{j}\frac{\delta}{\delta A_{j}}\big]}
		\int^{q_{f}}_{q_{i}}\mathcal{D}q\,e^{i\tilde{S}_{M}}\int^{\tilde{\mathcal{A}}_{j\,f}}_{{\tilde{\mathcal{A}}_{j\,i}}}\mathcal{D}
		\tilde{\mathcal{A}}_{\mu}\,e^{i\tilde{S}_A[\tilde{\mathcal{A}}]} \qquad
	\end{align}
	in which we write the answer in terms of the effective actions $\tilde{S}_M$
	and $\tilde{S}_A$, as in (\ref{eq:finalprop}).
	
	We can show the equivalence of this result to (\ref{eq:finalprop}) by noting
	that the remaining path-integral is independent of the longitudinal part of the
	gauge field. In the shift operator, the functional derivative
	$\partial_{j}\frac{\delta}{\delta A_{j}}$ then vanishes and we're left with an
	expression for the propagator $K_{fi}$ in the same form as (\ref{eq:finalprop})
	above, but with $\tilde{S}_C$ now written as
	\begin{equation}
		\tilde{S}_C \;=\;
		\int_{\partial\mathcal{V}}d^{3}x\,A_{k}\partial_{k}(\nabla^{-2}J^{0}),
		\label{tildeSC'}
	\end{equation}
	ie., as an integration by parts away from the expression for $\tilde{S}_C$
	in (\ref{tildeSC}). Again, we find that the charge is dressed by a Coulomb
	field.

	\subsection{Spinor Quantum Electrodynamics}
	\label{sec:diracQED}
	
	Here we give the derivation of the final result (\ref{eq:finalprop2}) for
	$K_{fi}$ starting from eqn. (\ref{Kfi-spin}) of the main text, ie., from
	\begin{eqnarray}
		K_{fi} &=&
		\int\mathcal{D}\Lambda\int_{\psi^{\Lambda_{i}}_{i}}^{\psi^{\Lambda_{f}}_{f}}\mathcal{D}\psi\mathcal{D}\bar{\psi}
		\int_{A^{\Lambda_{i}}_{\mu\,i}}^{A^{\Lambda_{f}}_{\mu\,f}}\mathcal{D}A_{\mu}  \;
		\Delta[A]\delta^{\mathcal{V}}\big(\mathcal{G}(A)\big)\delta^{\partial\mathcal{V}}\big(\mathcal{G}(A^{\Lambda})\big)e^{iS[A,\psi,\bar{\psi}]}
		\qquad
		\label{Kfi-spin2}
	\end{eqnarray}
	
	The $A_{0}$ integral in this formula can again be done without gauge fixing,
	and we can extract the transformations of the boundary data using
	exponentiations of the functional derivatives,
	\begin{align}
		&K_{fi} \;=\;
		\int\mathcal{D}\Lambda\,\delta^{\partial\mathcal{V}}\big(\mathcal{G}(A^{\Lambda})\big)
		e^{ \hat{\cal L}_{\Lambda}}
		\,\int_{\psi_{i}}^{\psi_{f}}\mathcal{D}\psi\mathcal{D}\bar{\psi}\int_{A_{\mu\,i}}^{A_{\mu\,f}}\mathcal{D}A_{\mu}\,\Delta[A]\delta^{\mathcal{V}}\big(\mathcal{G}(A)\big)e^{iS[A,\psi,\bar{\psi}]}.
	\end{align}
	in which the operator $\hat{\cal L}_{\Lambda}$ now takes the form
	\begin{eqnarray}
		\hat{\cal L}_{\Lambda} &=&
		\int_{\partial\mathcal{V}}d^{3}x\left[ie\psi\frac{\delta}{\delta\psi}+\partial_{\mu}\Lambda\frac{\delta}{\delta
			A_{\mu}}\right] \nonumber \\
		&=&
		\int_{\partial\mathcal{V}}d^{3}x\left[\Lambda\big(\partial_{j}\hat{E}^{j}-\hat{J}^{0}\big)-i\partial_{0}\Lambda\frac{\delta}{\delta
			A_{0}}\right]
		\label{hatL-D}
	\end{eqnarray}
	where the 2nd expression uses the relations (\ref{eq:fermionMomentum}),
	(\ref{eq:gaugefieldMomentum}).

	From this stage onwards, the manipulations are identical to those in the
	previous section except that the charge density in the boundary phase is an
	operator rather than a c-number. The resulting expression for the propagator is
	\begin{eqnarray}
		\label{eq:QEDpenultprop}
		K_{fi} &=& e^{i \hat{\cal L}_{\Lambda_{\cal G}}}
		\int_{\psi_{i}}^{\psi_{f}}\mathcal{D}\psi\mathcal{D}\bar{\psi}  \,
		\int_{A_{j\,i}}^{A_{j\,f}}\mathcal{D}A_{\mu}\,\Delta^{\mathcal{V}}[A]\delta^{\mathcal{V}}\big(\mathcal{G}(A)\big)e^{iS[A,\psi,\bar{\psi}]}
		\qquad
	\end{eqnarray}
	where now
	\begin{equation}
		\hat{\cal L}_{\Lambda_{\cal G}} \;=\;
		\int_{\partial\mathcal{V}}d^{3}x\Lambda_{\mathcal{G}}[A]\big[\partial_{j}\hat{E}^{j}-\hat{J}^{0}\big]
		\label{hatLG2}
	\end{equation}

	The final expression for the propagator will of course be independent of
	choice of $\mathcal{G}(A)$. For formal manipulations the most convenient choice
	is Coulomb gauge, because this sets the g-potential
	$\nabla^{-2}\partial^{j}A_{j}$ to zero, leaving only the invariant field
	components, and $\Lambda_{G}[A]=-\nabla^{-2}(\partial^{j}A_{j})$. Since this
	choice eliminates the dependence of the integral on the longitudinal part of
	$A_{j}$, the shift operator $\exp
	\big(i\int_{\partial\mathcal{V}}\Lambda\partial_{j}\hat{E}^{j}\big)$ will just
	give zero, and the propagator is then
	\begin{equation}
		\label{eq:finalprop2a}
		K_{fi} = e^{i\tilde{S}_C}
		\int_{\psi_{i}}^{\psi_{f}}\mathcal{D}\psi\mathcal{D}\bar{\psi} \,e^{i
			\tilde{S}_{M}}  \int^{\mathcal{A}_{j\,f}}_{{\mathcal{A}_{j\,i}}} \mathcal{D}
		\mathcal{A}_{j}\,e^{i \tilde{S}_A}.
	\end{equation}
	which is just eqn.(\ref{eq:finalprop2}) of the main text, as desired.
	
	Note that if we had chosen a different gauge fixing function
	$\mathcal{G}(A)$, the resulting gauge fixed action would look different, but
	this difference would only be temporary; the shift operator in (\ref{hatLG2})
	would no longer give zero in any other gauge, instead enacting a gauge
	transformation which would return the action to the form (\ref{tilde-S}), with
	the three terms given by eqns. (\ref{tildeSM})-(\ref{tildeSA}). In this form the
	theory is not manifestly Lorentz invariant, but this is simply because we
	evaluated $K_{fi}$ between two constant $t$ surfaces. In principle, one can
	chose a covariant gauge to compute the path-integral as long as one also
	evaluates the necessary shift of the longitudinal mode in the final expression.



\begin{thebibliography}{99}
		
		
		\bibitem{bergmann}   See, eg., R. Utiyama, Phys. Rev. {\bf 101}, 1597
		(1956); T.W.B. Kibble, J. Math. Phys. {\bf 2}, 212 (1960); P.G. Bergmann, Phys.
		Rev. {\bf 124}, 274 (1961); and J.D. Norton, Rep. Prog. Phys. 56, 791 (1993) for
		a survey of some of the earlier discussion.
		
		\bibitem{kuchar}     K. Kuchar, in ``{\it Proceedings of the 4th
			Canadian
			Conference on General Relativity and Relativistic Astrophysics}",
		ed. G. Kunstatter et al. (World Scientific, 1992).
		
		\bibitem{woodard}    N. C. Tsamis, R. P. Woodard, Ann. Phys. (N.Y.) {\bf
			215},
		96 (1992).
		
		\bibitem{marolf06}   S. B. Giddings, D. Marolf, J.B. Hartle, Phys. Rev.
		{\bf D 74}, 064018 (2006)
		
		\bibitem{hartle88}   J.B. Hartle, Phys. Rev. {\bf D37}, 2818 (1988);
		Phys. Rev. {\bf D38}, 2985 (1988)
		
		\bibitem{hartle94}   J.B. Hartle, Phys. Rev. {\bf D49}, 6543.
		
		\bibitem{oeckl}      R. Oeckl, Phys. Lett. {\bf B575}, 318 (2003); and
		Adv. Theor. Math. Phys. {\bf 2}, 451 (2015).
		
		\bibitem{rovelliB}   F. Conrady et al., Phys. Rev. {\bf D 69}, 064019
		(2004); R. Oeckl, Phys. Rev. {\bf D 73}, 065017 (2006).
		
		\bibitem{hawkingPI}  G.W. Gibbons, S.W. Hawking, Phys. Rev. {\bf D15},
		2752 (1977); and S.W. Hawking, pp. 746-789 in ``{\it General Relativity: An
			Einstein Centenary Survey}", ed. S.W. Hawking, W. Israel (Cambridge Univ. Press,
		1979)
		
		\bibitem{hartleH}    J.B. Hartle, S.W. Hawking, Phys. Rev. {\bf D28},
		2960 (1983)
		
		\bibitem{halliwell} J.J. Halliwell and J.B. Hartle, Phys. Rev. {\bf
			D43}, 1170 (1991)
		
		\bibitem{Qcosmo}     For an introduction to quantum cosmology, see, eg.,
		C. Kiefer, ``{\it Quantum Gravity}" (Oxford Univ. press, 2012)
		
		\bibitem{kibble89}   T.W.B. Kibble, Comm. Math. Phys. {\bf 64}, 73
		(1978); T.W.B. Kibble, pp. 63-80 in ``{\it Quantum Gravity 2}\,'', ed. C.J.
		Isham, R. Penrose, D.W. Sciama (O.U.P., 1981)
		
		\bibitem{weinberg89} S. Weinberg, Phys. Rev. Lett. {\bf 62}, 485 (1989);
		S. Weinberg, Ann. Phys. (NY) {\bf 194}, 336 (1989)
		
		\bibitem{kibbleQG}   T.W.B. Kibble, S. Randjbar-Daemi, J. Phys A{\bf
			13}, 141 (1980)
		
		\bibitem{penrose96}  R. Penrose, Gen. Rel. Grav. {\bf 28}, 581 (1996);
		R. Penrose, Phil. Trans. Roy. Soc. Lond. {\bf A356}, 1927 (1998); W Marshall, C
		Simon, R Penrose, D Bouwmeester,  Phys. Rev. Lett.
		{\bf 91}, 130401 (2003). See also D. Kleckner et al., New J. Phys. {\bf
			10}, 095020 (2008)
		
		\bibitem{bassi14}    A. Bassi et al., Rev. Mod. Phys. {\bf 85}, 471
		(2013)
		
		\bibitem{stampCWL}   P.C.E. Stamp, New J. Phys. {\bf 17} 065017 (2015);
		A.O. Barvinsky, D. Carney, and P. Stamp, Phys. Rev, {\bf D 98}, 084052 (2018).
		
		\bibitem{CWL2}       A.O. Barvinsky, J. Wilson-Gerow, P.C.E. Stamp,
		Phys. Rev. D{\bf 103}, 064028 (2021)
		
		\bibitem{CWL3}       J. Wilson-Gerow, P.C.E. Stamp, /arXiv 2011.14242
		
		
		
		
		
		
		
		
		
		\bibitem{morette82}  C. Morette-DeWitt, Comm. Math. Phys. {\bf 28}, 47
		(1972).
		
		
		\bibitem{AhB57}      Y. Aharonov, D. Bohm, Phys. Rev. {\bf 115}, 485
		(1959); Y. Aharonov, D. Bohm, Phys. Rev. {\bf 123}, 1511 (1961)
		
		\bibitem{no-int}     PS Epstein, Am. J. Phys. {\bf 13}, 127 (1945); M
		Renninger, Z. Physik {\bf 158}, 417 (1960); RH Dicke, Am. J. Phys. {\bf 49}, 925
		(1981).
		
		
		\bibitem{fracStat}   M.G.G. Laidlaw, C.Morette-DeWitt, Phys. Rev. {\bf
			D3}, 1375 (1971); J.M. Leinaas, J. Myrheim, Nuovo Cim. {\bf 37B}, 1 (1977)
		
		\bibitem{wu84}       Y-S Wu, Phys. Rev. Lett. {\bf 52}, 2103 (1984)
		
		\bibitem{stone}      M. Stone, ``{\it Quantum Hall Effect}" (World
		Scientific, 1992)
		
		
		
		
		
		
		
		\bibitem{thorneCTC}  G. Klinkhammer, K.S. Thorne, unpublished; and D.S.
		Goldwirth, M.J. Perry, T. Piran, K.S. Thorne, Phys. Rev. {\bf D49}, 3951 (1994).
		
		
		\bibitem{visser}     M. Visser, ``{\it Lorentzian Wormholes}" (Springer
		Verlag, 1996).
		
		
		\bibitem{coleman91}    S. Coleman, J.B. Hartle, T. Piran, S. Weinberg
		(eds.) ``{\it Quantum cosmology and baby universes},” in Proceedings, 7th
		Jerusalem Winter School for Theoretical Physics (Singapore: World Scientific,
		1991), and refs. therein.
		
		\bibitem{hebeker18}     A. Hebeecker, T. Mikhail, P. Soler, Front.
		Astron. Space Sci. {\bf 5}, 35 (2018), and refs. therein.
		
		
		
		
		\bibitem{penington2020} G. Penington, S.H. Shenker, D. Stanford, and Z.
		Yang, arxiv:1911.11977 [hep-th]
		
		\bibitem{almheiri2020} A. Almheiri, T. Hartman, J. Maldacena, et al., J.
		High Energ. Phys. {\bf 2020,} 13 (2020)
		
		\bibitem{vanraamsdonk2020} M. Van Raamsdonk, arXiv:2008.02259 [hep-th]
		
		\bibitem{marolf2020} D. Marolf, H. Maxfield,  J. High Energ. Phys. {\bf
			2020,} 44 (2020)
		
		\bibitem{giddings2020} S.B. Giddings, G.J. Turiaci,  J. High Energ.
		Phys. {\bf 2020,} 194 (2020)
		
		\bibitem{turok}     J. Feldbrugge, J.-L. Lehners, N. Turok, Phys. Rev.
		Lett. {\bf 119}, 171301 (2017), and Phys. Rev. D{\bf 95}, 103508 (2017); A. Di
		Tucci, J. Feldbrugge, J.-L. Lehners, N. Turok, Phys. Rev. D{\bf 100}, 063517
		(2019); see also A. Baldazzi, R. Percacci, V. Skrinjar, Class. Q. Grav. {\bf
			36}, 105008 (2019).
		
		\bibitem{vilenkin}  A. Vilenkin, M. Yamada, Phys. Rev. D{\bf 98}, 066003
		(2018), and Phys. Rev. D{\bf 99}, 066010 (2019); see also A. Di Tucci, J.-L.
		Lehners, Phys. Rev. Lett. {\bf 122}, 201302 (2019).
		
		\bibitem{bojowald}  M. Bojowald, S. Brahma, Phys. Rev. Lett. {\bf 121},
		201301 (2018), and Phys. Rev. D{\bf 102}, 106023 (2020).
		
		
		
		
		\bibitem{dirac50}    P.A.M. Dirac, Can. J. Math. {\bf 2}, 129 (1950)
		
		\bibitem{dirac64}    P.A.M. Dirac, ``{\it Lectures on Quantum
			Mechanics}" (Belfer graduate school of Science, Yeshiva Univ, New York, 1964)
		
		\bibitem{dirac55}    P.A.M. Dirac, Can. J. Phys. {\bf 33}, 650 (1955)
		
		\bibitem{deWitt67c}  B.S. DeWitt, Phys. Rev. {\bf 160}, 1113 (1967)
		
		
		
		\bibitem{ashtekar}   A. Ashtekar, ``{\it Asymptotic Quantization}"
		(Bibliopolis, 1987)
		
		\bibitem{strominger} A. Strominger, ``{\it Lectures on the Infrared
			Structure of Gravity and Gauge Theory}" (Princeton Univ. Press, 2018)
		
		\bibitem{CQG18}      J. Wilson-Gerow, C. DeLisle, P.C.E. Stamp,  Class.
		Q. Grav. {\bf 35}, 164001 (2018); see also C. DeLisle, J. Wilson-Gerow, P.C.E.
		Stamp, /ArXiv: 1905.05333
		
		\bibitem{vedral17}   C. Marletto, V. Vedral, Phys. Rev. Lett. {\bf 119},
		240402 (2017)
		
		\bibitem{milburn17}  S. Bose et al., Phys. Rev. Lett. {\bf 119}, 240401
		(2017)
		
		\bibitem{beilok18}   C. Anastopoulos, B.-L. Hu, /arXiv 1804.11315.
		
		\bibitem{belenchia18}  A. Belenchia, R. M. Wald, F. Giacomini, E.
		Castro-Ruiz, C. Brukner, M. Aspelmeyer, Phys. Rev. D{\bf 98}, 126009 (2018)
		
		\bibitem{anasHu}     C. Anastopoulos, B.-L. Hu,  Class. Quantum Grav.
		{\bf 32}, 165022 (2015)
		
		\bibitem{onigaW}     T. Oniga, C.H.-T. Wang, Phys. Rev. {\bf D 93},
		044027 (2016)
		
		\bibitem{blencowe}   M.P. Blencowe, Phys. Rev. Lett. {\bf 111}, 021302
		(2013)
		
		\bibitem{jordanMSc}  J. Wilson-Gerow, MSc thesis (UBC, Sept. 2017)
		
		\bibitem{FP67}       L.D. Faddeev, V.N. Popov, Phys. Lett. {\bf B25}, 29
		(1967)
		
		\bibitem{gribov}     V.N. Gribov, Nucl. Phys. {\bf B139}, 1 (1978); a
		more recent
		review is in N. Vandersickel, D. Zwanziger, Phys. Rep. {\bf 520}, 175
		(2012)
		
		\bibitem{mandelstam62} S. Mandelstam, Ann. Phys. {\bf 19}, 1 (1962), and
		{\it ibid.}, 25 (1962)
		
		\bibitem{shabanov}   S.V. Shabanov, J. Phys. {\bf A23}, 3245 (1990);
		L.V. Prokhorov, D.V. Fursaev, S.V. Shabanov, Theor. Math. Phys. (USSR), {\bf
			97}, 1323 (1993)
		
		\bibitem{giddings}   W. Donnelly, S.B. Giddings, Phys. Rev.
		{\bf D 93}, 024030 (2016)
		
		\bibitem{fradkin}    E.S. Fradkin, Nucl. Phys. {\bf 76}, 588 (1966)
		
		\bibitem{fried}      H.M. Fried, ``{\it Green's Functions and Ordered
			Exponentials}" (Cambridge Univ. Press, 2002)
		
		\bibitem{gervais}    J.-L. Gervais, D. Zwanziger, Phys. Lett. {\bf 94B},
		389 (1980)
		
		\bibitem{carney}     D. Carney, L. Chaurette, D. Neuenfeld, G.W.
		Semenoff, Phys. Rev. {\bf D 97}, 025007 (2018)
		
		
		\bibitem{jackson}    J.D. Jackson, ``{\it Classical Electrodynamics}"
		(Wiley, 1962)
		
		\bibitem{wilson20}   J. Wilson-Gerow, to be published.
		
		\bibitem{testa}      G.C. Rossi, M. Testa, Nucl. Phys. {\bf B 163},
		109 (1980); and Nucl. Phys. {\bf B 176}, 477 (1980); and F. Mattei, C.
		Rovelli, S. Speziale, M. Testa, Nucl. Phys. {\bf B 739}, 234 (2006);
		
		
		
		
		
		\bibitem{colby}      See C. Delisle, to be published. We thank C.
		Delisle for extensive discussion on this point.
		
		\bibitem{lee-wald}   J. Lee, R.M. Wald, J. Math. Phys. {\bf 31}, 725
		(1990)
		
		\bibitem{BondiS}     H. Bondi, A.W.K. Metzner and M.I.G. Van der Berg,
		Proc. Roy. Soc. {\bf A 269}, 21 (1962); R.K. Sachs, Proc. Roy. Soc. {\bf A 270},
		103 (1962), and Phys. Rev. {\bf 128}, 2851 (1962)
		
		\bibitem{largeGT}    The connection between large gauge transformations
		and soft theorems was first discussed for QED by R. Ferrari, L.E. Picasso, Nucl.
		Phys. {\bf B31}, 316 (1971). For more general recent discussions, see, eg., Y.
		Hamada, M.-S. Sheo, G. Shiu, Phys. Rev. {\bf D 96}, 105013 (2017); Y. Hamada, G.
		Shiu, Phys. Rev. Lett. {\bf 120}, 201601 (2018); and G. Compere, Phys. Rev.
		Lett. {\bf 123}, 021101 (2019).
		
		\bibitem{IRstuff}    For a thorough review of soft photons and gravitons
		and the connection with conservation laws, see \cite{strominger}. For the
		connection with memory effects, see, eg., P. Mao, W.-D. Tan, Phys. Rev. {\bf D
			101}, 124015 (2020), and refs. therein.
		
		\bibitem{ashtekartalk} A. Ashtekar, \textit{The BMS group, conservation
			laws, and soft gravitons}, Talk presented at the \textit{Perimeter Institute for
			Theoretical Physics}, 8 November 2016,
		\href{https://pirsa.org/16080055/}{https://pirsa.org/16080055/}
		
		\bibitem{ashtekarhansen} A. Ashtekar and R.O. Hansen, J. Math. Phys.
		\textbf{19}, 1542 (1978)
		
		\bibitem{prabhupapers} K. Prabhu, J. High Energ. Phys. \textbf{2018},
		113 (2018); V. Chandrasekaran and K. Prabhu, J. High Energ. Phys. \textbf{2019},
		229 (2019); K. Prabhu, J. High Energy. Phys. \textbf{2019}, 148 (2019); K.
		Prabhu and I. Shehzad 2020 Class. Quantum Grav. \textbf{37} 165008; K. Prabhu
		and I. Shehzad, arXiv:2110.04900 [gr-qc]
		
		\bibitem{kibbleIR}   T.W.B. Kibble, J. Math. Phys. {\bf 9}, 315 (1968);
		Phys. Rev.
		{\bf 173}, 1527 (1968), Phys. Rev. {\bf 174}, 1882 (1968); and Phys.
		Rev.
		{\bf 175}, 1624 (1968)
		
		\bibitem{chung}      V. Chung, Phys. Rev. {\bf 140}, B1110 (1965)
		
		\bibitem{faddeevK}   P.P. Kulish, L.D. Faddeev, Theor. Math. Phys. {\bf
			4}, 745 (1970).
		
		\bibitem{newsymmetries2014} T. He, P. Mitra, A.P. Porfyriadis, A.
		Strominger, J. High Energ. Phys. \textbf{2014}, 112 (2014)
		
		\bibitem{kapec2017} D. Kapec, M. Perry, A.-M. Raclariu, A. Strominger,
		Phys. Rev. \textbf{D 96}, 085002 (2017)
		
		\bibitem{akhoury2018} S. Choi, R. Akhoury, J. High Energ. Phys.
		\textbf{2018}, 171 (2018)
		
		\bibitem{strominger2014} A. Strominger, J. High Energ. Phys.
		\textbf{2014}, 152 (2014)
		
		\bibitem{low1958} F.E. Low, Phys. Rev. \textbf{110}, 974 (1958)
		
		\bibitem{Jordan-PCES}  J. Wilson-Gerow, P.C.E. Stamp, to be published;
		and see J. Wilson-Gerow, PhD thesis (Univ. of British Columbia, Sep. 2021)
		
		
		
		
		
		
		
		
		
		
	\end{thebibliography}
\end{document}